\documentclass[%
 reprint,
 superscriptaddress,
preprintnumbers,
nofootinbib,
 amsmath,amssymb,
 aps,
]{revtex4-1}
\usepackage{amsmath}
\usepackage{amssymb}
\usepackage{amsthm}
\usepackage{MnSymbol}
\usepackage{nicefrac}
\usepackage{flushend}
\usepackage{amsfonts}
\usepackage{dsfont}

\usepackage[markup=nocolor]{changes}

\usepackage{graphicx}
\usepackage{tabularx}

\usepackage[normalem]{ulem}

\usepackage{dcolumn}
\usepackage{bm}

\usepackage{xcolor}
\usepackage{slashed}
\usepackage{empheq}
\usepackage{tensor}
\usepackage{hyperref}
\newcommand{\bea}{\begin{eqnarray}}
\newcommand{\eea}{\end{eqnarray}}
\newcommand{\be}{\begin{equation}}
\newcommand{\ee}{\end{equation}}

\definecolor{darkgreen}{rgb}{0,0.4,0}

\newcolumntype{N}{@{}m{0pt}@{}}
\newcolumntype{M}[1]{>{\centering\arraybackslash}m{#1}}
\begin{document}
\author{Astrid Eichhorn} \email{eichhorn@sdu.dk}
\affiliation{CP3-Origins, University of Southern Denmark, Campusvej
  55, DK-5230 Odense M, Denmark} 
  \affiliation{Institut f\"{u}r
  Theoretische Physik, Universit\"{a}t Heidelberg, Philosophenweg 16,
  69120 Heidelberg, Germany}

\author{Alessia Platania} \email{platania@thphys.uni-heidelberg.de}
\affiliation{Institut
 f\"{u}r Theoretische Physik, Universit\"{a}t Heidelberg,
 Philosophenweg 16, 69120 Heidelberg, Germany}
 
\author{Marc Schiffer} \email{schiffer@thphys.uni-heidelberg.de}
\affiliation{Institut
 f\"{u}r Theoretische Physik, Universit\"{a}t Heidelberg,
 Philosophenweg 16, 69120 Heidelberg, Germany}
 
\title{Lorentz invariance violations 
in the interplay of quantum gravity with matter}

\begin{abstract}
{We explore the interplay of matter with quantum gravity with a preferred frame to highlight that the matter sector cannot be protected from the symmetry-breaking effects in the gravitational sector. Focusing on Abelian gauge fields, we show that quantum gravitational radiative corrections induce Lorentz-invariance-violating couplings for the Abelian gauge field. In particular, we discuss how such a mechanism could result in the possibility to translate observational constraints on Lorentz violation in the matter sector into strong constraints on the Lorentz-violating gravitational couplings. }
\end{abstract}

\maketitle
\section{Introduction}
In our current theoretical framework for the building blocks of nature, symmetries play a central role, not least because of the visionary insights of the mathematician Emmy Noether. As we push our understanding of the fundamental interactions of nature to smaller distance scales (higher energy scales), the fate of symmetries across different scales is a key piece of information. In particular, proposals that Lorentz symmetry is an emergent symmetry at low energies, but broken at and beyond the Planck scale, have been made, e.g., in \cite{Gambini:1998it,Jacobson:2000xp,Carroll:2001ws,Magueijo:2001cr,Magueijo:2002am,GrootNibbelink:2004za,Horava:2009uw,Liberati:2009pf,Kharuk:2015wga}, concertedly with suggestions for experimental tests, e.g., \cite{AmelinoCamelia:1997gz}, see also the reviews \cite{Mattingly:2005re,AmelinoCamelia:2008qg,Hossenfelder:2012jw,Liberati:2013xla}. In a quantum gravitational context, this implies, e.g., the breakdown of diffeomorphism invariance to foliation-preserving diffeomorphism symmetry, due to the presence of a preferred frame. The remaining theory is therefore invariant under three-dimensional rotations on the spatial slices orthogonal to the time-like vector $n_{\mu}$. Scenarios, where diffeomorphism invariance is broken due to the existence of more general background structures and therefore without residual foliation preserving diffeomorphism invariance, are investigated, for example in \cite{Sudarsky:2005ab,Cohen:2006ky,Bailey:2006fd,Bernardini:2007ez,Ackerman:2007nb,Tasson:2014dfa}.
The observation of gravitational waves from a binary neutron star merger~\cite{TheLIGOScientific:2017qsa,GBM:2017lvd,Monitor:2017mdv}, as well as from binary black hole mergers \cite{Abbott:2016blz,Abbott:2016nmj,Abbott:2018lct}, has opened up novel observational opportunities in this area, cf.~\cite{Gumrukcuoglu:2017ijh,Mewes:2019dhj}, and \cite{Yunes:2016jcc,Ramos:2018oku}, respectively. Yet, most experimental constraints in this field come from the non-observation of Lorentz-symmetry violation in the matter sector, see, e.g., \cite{Wolf:2004gg,Antonini:2005yb,AmelinoCamelia:2005qa,Wolf:2006uu,Kostelecky:2006ta,Muller:2007es,5ecad05bdf014116b23a59df9fda1c5d,Chung:2009rm,Michimura:2013kca,Nagel:2014aga,Tasson:2014dfa,Lo:2014yea,Chen:2016eli,Kostelecky:2016kkn,Bars:2016mew,Fu:2016fmf,Wiens:2016bqh,Lai:2017bbl,Lehnert:2018lce,Sanner:2018atx,Goryachev:2018nln,Megidish:2018sey,Kelly:2018zbx,Ding:2019aox} and references therein, and \cite{Kostelecky:2008ts} for a summary of experimental bounds. For searches of Lorentz-symmetry violations in the pure gravitational sector, see, e.g., \cite{Kostelecky:2015dpa,Hees:2016lyw,Mewes:2019dhj}.
In fact, we expect that Lorentz-invariance violation (LIV) cannot occur just in the matter or the gravitational sector without also percolating into the respective other sector. This is due to the simple fact that any form of matter gravitates, and therefore the interaction between the two sectors cannot be switched off. Typically, loop corrections in such coupled systems result in the impossibility to isolate violations of symmetry to just one sector. This argument has been made, e.g., in~\cite{Kostelecky:2003fs,Collins:2004bp,Kostelecky:2010ze,Pospelov:2010mp,Liberati:2012jf,Belenchia:2016lfc,Bluhm:2019hct}.
Here, we support the general argument by an explicit calculation that provides an example showing that indeed Lorentz-symmetry violation in quantum gravity necessarily percolates into the matter sector. In particular, we show that under certain assumptions that we will spell out in detail below, the 
``amount" of Lorentz-symmetry violation in the quantum-gravity sector (measured by the deviation of dimensionless couplings singling out a preferred frame from zero) correlates with the amount of Lorentz-symmetry violation in the matter sector. Typically, gravitational couplings of $\mathcal{O}(10^{-n})$ induce LIV-matter couplings of about the same order. 
Hence, strong constraints on LIV couplings in the matter sector typically imply similarly strong constraints on the gravitational LIV couplings. Moreover, we highlight that the induced LIV couplings in the matter sector include \emph{marginal} couplings, which are -- unlikely their Planck-scale suppressed counterparts -- observationally easier to constrain.

We will work in a model of quantum gravity with foliation-preserving diffeomorphism symmetry, coupled to a single gauge field. This serves as a toy model of the Abelian gauge sector of the Standard Model (corresponding to electromagnetism below the scale of spontaneous electroweak symmetry breaking), in which the presence of LIV-couplings is observationally strongly constrained by astrophysical observations as well as laboratory experiments.

We will perform a Renormalization Group (RG) study of the system.
In summary, we will show that the following hold within our toy model and within the technical limitations of our study, to be discussed below: i) quantum gravitational dynamics which single out a preferred frame necessarily generate Lorentz-invariance violations for matter in the ultraviolet (UV) and ii) under appropriate conditions, this violation must necessarily persist in the infrared. In other words, the violation of Lorentz symmetry in the gravitational sector percolates into the matter sector in the UV. Under appropriate conditions -- amounting to the existence of a  infrared (IR) attractive fixed point in the RG flow of the matter coupling -- this symmetry violation must persist into the infrared, where it is accessible to experimental tests. The qualitative picture is therefore that small violations of Lorentz invariance in the UV will in general grow towards the IR. The existence of an IR attractive fixed point prevents the violations of Lorentz invariance 
from growing even larger.  Experimental data provide strong constraints on LIV in  realistic models which include all relevant degrees of freedom. Due to the points i) and ii), these experimental bounds can in turn be used to put constraints on  UV violations of Lorentz symmetry in the gravity sector. 
 
Further, we argue that the various terms  in the Standard Model Extension (SME) \cite{Colladay:1998fq,Kostelecky:2002hh}, see \cite{Bluhm:2005uj,Tasson:2014dfa} for reviews, are typically not independent when derived from an underlying microscopic model. A given microscopic model (defined by a set of values for the gravitational couplings) most likely generates \emph{all} terms in the SME with a given set of symmetries (e.g., CPT symmetry might or might not hold in a given quantum-gravity setting). Typically, we expect that all these couplings are generated with dimensionless counterparts of order~$1$, if the gravity couplings are order~$1$. 
This is of course a standard naturalness argument -- a given microscopic model might circumvent this and in fact provide an explanation why a set of couplings is ``unnaturally" small. Here, we will provide one example to show that a given set of microscopic gravitational couplings typically generates the strongly constrained marginal LIV couplings together with the less strongly constrained higher-order couplings. Therefore, weaker direct experimental constraints on the higher-order couplings could, under certain conditions, actually be supplemented by \emph{indirect} strong consistency constraints.

Let us stress that we perform our study within a toy model that does not account for the existence of Standard Model degrees of freedom beyond the Abelian gauge field, and that does not account for the difference between the Abelian hypercharge field at high energies and the photon at low energies, which is due to electroweak symmetry breaking. Yet, we do not expect that these additional intricacies can impact the main outcomes of our study, at least at the qualitative level. The interplay of electroweak symmetry breaking with LIV has been explored in \cite{Colladay:1998fq}.

This paper is organized as follows. In Sec.~\ref{sec: LIVimp} we introduce the system of an Abelian gauge field coupled to foliation-preserving-diffeomorphism invariant gravity, including the foliation structure and a LIV term for the Abelian gauge field. In Sec.~\ref{sec: GravMatInterplay}, we investigate the impact of Lorentz invariance violations in the gravity sector onto the Abelian gauge field, and discuss the role of (pseudo) fixed points as attractors and repulsors in the RG flow. In Sec.~\ref{sec: constraints} we study the regions in the  gravitational parameter space giving rise to a universal value for the matter LIV coupling $\zeta$ at the Planck scale using the flow equation obtained in our approximation. 
Further, we aim at highlighting the constraining power that arises from the type of study we perform here. For this purpose, we use experimental constraints on Lorentz symmetry violations in the photon sector. By imposing these bounds on the LIV coupling for the Abelian gauge field in our toy model, we arrive at strong constrains on the gravity LIV couplings. We stress that these constraints are subject to the systematic uncertainties of our study, and the difference of our toy model to the full SME coupled to gravity. Therefore, these constraints cannot yet be viewed on the same footing as direct experimental constraints on the gravity LIV couplings. However, our study clearly highlights the potential constraining power of the gravity-matter interplay within a LIV setting. This strongly motivates upgraded studies which go beyond our toy model, in order to bring the power of this idea to bear on quantum gravity. We understand our present study as a blueprint that exemplifies this idea, see also the corresponding comments in \cite{Gumrukcuoglu:2017ijh}.
In Section~\ref{sec: HigherOrders} we provide an explicit example to highlight that any marginal LIV coupling is likely to be gravitationally induced concertedly with a higher dimensional LIV couplings. Their dimensionless couplings are typically of the same order also due to the direct interplay between them. This could give rise to indirect constraints on higher order LIV couplings, which are expected to  be  stronger than direct experimental constraints. Finally, in Sec.~\ref{sec: Concl} we summarize our results and provide a brief outlook on future perspectives.

\section{Impact of quantum gravity with a preferred frame on matter}
\label{sec: LIVimp}
To investigate how the breaking of Lorentz symmetry in the matter sector is influenced by symmetry-breaking terms in the gravitational sector, we will explore the Wilsonian scale dependence of a matter LIV coupling. For this study, we make use of the well suited tool of the  functional Renormalization Group (RG) (see, e.g., \cite{Berges:2000ew,Pawlowski:2005xe,Gies:2006wv,Rosten:2010vm} for reviews). Due to a suitable infrared (IR) and UV regularization, it  implements the Wilsonian idea of momentum-shell wise integration of quantum fluctuations and allows to investigate the scale dependence of quantum field theories within and beyond perturbation theory.
More specifically, the functional RG relies on a flow equation, the Wetterich equation \cite{Wetterich:1992yh, Ellwanger:1993mw,Morris:1993qb,Reuter:1993kw,Tetradis:1993ts}, that is a functional integro-differential equation for the scale-dependent effective action~$\Gamma_k$.  The latter provides the RG scale $k$ dependent  
equations of motion for the expectation values of the quantum fields. In the limit~$k \rightarrow \infty$, $\Gamma_k$ essentially provides the microscopic or classical action, whereas in the physical limit~$k \rightarrow 0$ all quantum fluctuations are included, and~$\Gamma_k$ reduces to the standard effective action. The Wetterich equation gives rise to flow equations for the couplings, which encode how the couplings in the effective dynamics change, as quantum fluctuations with momenta of the order $k$ are integrated over. The functional RG is applied in a broad range of contexts; selected examples in models with interacting fixed points include the $O(N)$ model, e.g., \cite{Canet:2003qd,Litim:2010tt,Juttner:2017cpr,Balog:2019rrg} and the Gross-Neveu(-Yukawa) model \cite{Braun:2010tt,Classen:2015mar,Knorr:2016sfs}. In all cases, quantitative agreement with other methods was achieved by extending the truncation according to the canonical power-counting of higher-dimensional operators.
For more details on the functional RG for the present setup, see Appendix~\ref{sec: FRG}. For other ideas to constrain physics beyond the Standard Model using the functional RG in the context of Lorentz invariant asymptotically safe gravity, see, e.g., \cite{deBrito:2019epw,Reichert:2019car}.
Let us stress that the functional RG relies on using the Euclidean four-momentum, and therefore provides access to the scale-dependence in Riemannian quantum-gravity settings. Performing a continuation of the results to a Lorentzian setting in the presence of dynamical gravity is an outstanding challenge. Thus, we work under the assumption that our results carry over to a Lorentzian setting. 

To explore the consequences of the existence of a preferred frame, we adapt our setup to allow direct access to the foliation structure of the manifold \cite{Knorr:2018fdu}, $\mathcal{M} = \Sigma \times \mathbb{R}$, where $\Sigma$ is a Riemannian 3-manifold and $\mathbb{R}$ is the ``time'' direction (i.e., a preferred spatial direction in our Euclidean spacetimes).
In this setup, the full metric~$g_{\mu\nu}$ is expressed in terms of a tensor~$\sigma_{\mu\nu}$, encoding the three-metric in $\Sigma$ in a covariant way, and an orthogonal, normalized time-like vector $n_{\mu}$, i.e.,
\bea
g_{\mu\nu}&=&\sigma_{\mu\nu}+n_{\mu}n_{\nu}\,\notag\\
g^{\mu\nu}n_{\mu}\sigma_{\nu\rho}&=&0,\notag\\
g^{\mu\nu}n_{\mu}n_{\nu}&=&1.
\label{eq: cond}
\eea
We refer the reader to Appendix~\ref{sec: Foliation} for details. 
The time-like vector~$n_{\mu}$ can be used to single out a preferred frame, such that full diffeomorphism invariance is broken to foliation-preserving diffeomorphisms.  As we use the functional RG, we choose the four-metric to be a Riemannian metric. The vector field $n_{\mu}$ singles out  a distinguished direction in which an analytic continuation of the metric could be performed and in this sense it singles out a time direction.  Using the decomposition~\eqref{eq: cond}, quantum fluctuations of the full metric~$g_{\mu\nu}$ can be expressed in terms of fluctuations in~$\sigma_{\mu\nu}$ and $n_\mu$. In the following, we will denote the expectation values of these two fields simply by~$\sigma_{\mu\nu}$ and~$n_{\mu}$.

In our approximation, we will parameterize the dynamics of the diffeomorphism invariant sector of gravity via the Einstein-Hilbert action with the scale-dependent Newton coupling $G_{\rm N} (k)$ and the cosmological constant $\Lambda(k)$,
\begin{equation}
\Gamma_{k}^{\rm EH}
=\frac{1}{16\pi G_{\rm N}(k)}\int\! \sqrt{\det(g_{\rho\sigma})}\,\,(-R+2\Lambda(k)).
\label{eq: EH}
\end{equation}
We use $\Gamma_k$ to indicate that this is an ansatz for the scale-dependent effective action entering the Wetterich equation.
Additionally, we include in our truncation of the dynamics the effect of all the canonically most relevant operators that break Lorentz invariance. The three independent \cite{Knorr:2018fdu} tensor structures containing up to two derivatives are given by\footnote{The couplings associated with the breaking of Lorentz symmetry in the gravitational sector in \cite{Gumrukcuoglu:2017ijh} and in our work are related via $\gamma=k_0$, $\beta=k_2$ and $\alpha=-a_1$.}
\bea
\Gamma_{k}^{\rm Grav, \,LIV}
&=&\frac{1}{16\pi G_{\rm N}(k)}\int\! \sqrt{\det(g_{\rho\sigma})}\,\,\Bigl(k_2(k) K^{\mu\nu}K_{\mu\nu}\nonumber\\
&{}&\quad \quad+k_0 (k)K^2+a_1 
(k)\mathcal{A}^{\mu}\mathcal{A}_{\mu}\Bigr),
\label{eq: breakingaction}
\eea
with symmetry-breaking and scale-dependent couplings $k_2(k),\,k_0(k)$ and $a_1(k)$. Here, the extrinsic curvature on spatial slices $K_{\mu\nu}$ is orthogonal to the normal vector
\begin{equation}
\label{eq: extrorth}
n^{\mu}K_{\mu\nu}=0.\,
\end{equation}
In terms of the fields~$n_{\mu}$ and~$\sigma_{\mu\nu}$, it reads
\begin{equation}
K_{\mu\nu}=\frac{1}{2}(n^{\alpha}D_{\alpha}\sigma_{\mu\nu}+D_{\mu}n_{\nu}+D_{\nu}n_{\mu}).
\end{equation}
In addition, $K$ is the trace of the extrinsic curvature and~$\mathcal{A}_{\mu}$ is the acceleration vector
\begin{equation}
\mathcal{A}_{\mu}=n^{\alpha}D_{\alpha}n_{\mu}.
\end{equation}
All the ``breaking terms'' are invariant under foliation-preserving diffeomorphisms but not under full diffeomorphisms, thus singling out a physical, preferred frame.
All other terms with these symmetries and at second order in derivatives are, up to a total derivative which is neglected, related via the Gauss-Codazzi equations. Therefore, the truncation we consider in this paper corresponds to the IR limit of Horava-Lifshitz gravity coupled to an Abelian gauge field, which is perturbatively renormalizable. Due to the Wilsonian treatment, additional operators, such as Lorentz-invariance violating matter-gravity operators, are induced at lower scales. Besides that, the Wilsonian treatment allows for a broader perspective, since it allows the study of theories, where some other theory sets in beyond a cutoff scale in the far UV. The gravitational part of the action is complemented with the standard gauge-fixing and ghost term, and a constraint term which implements the foliation structure of the system. Following \cite{Knorr:2018fdu}, we implement the latter like a gauge condition into the path integral. For details on the implementation of the foliation constraint, see App.~\ref{sec: Foliation}. Since the conditions~\eqref{eq:  cond} are second-class constraints, as opposed to gauge conditions which are first-class constraints, their implementation might require a modification of this procedure~\cite{Eichhorn2019}. This contributes to the systematic uncertainty of our results, which are however expected to be dominated by truncation errors.  This ansatz for $\Gamma_k$ is based on canonical power counting, i.e., the truncation of the theory space is chosen by including operators by canonical relevance. Such a truncation is expected to reliably capture physics in the perturbative regime, where higher-order couplings typically remain small and irrelevant. In fact, even in a setting with interacting fixed points, such truncations could be reliable.
Indeed, in the context of diffeomorphism invariant gravity, there are indications that the asymptotically safe fixed point lies in a near-perturbative regime \cite{Denz:2016qks,Eichhorn:2018akn,Eichhorn:2018ydy,Eichhorn:2018nda}, where higher order operators follow their canonical scaling \cite{Falls:2013bv,Falls:2014tra,Falls:2017lst, Falls:2018ylp}.  For a non-gravitational example showing the  the convergence of a truncation based on canonical power counting for an interacting fixed point, see, e.g.,  \cite{Balog:2019rrg}. 

\begin{table*}[ht]
	\begin{tabular}{|M{1.1cm}|c|c|l|N}
		\textbf{Bound} & \textbf{Year} & \textbf{Ref.} & \textbf{Method} \\\hline\hline
		$10^{-37}$ & $2006$& \cite{Kostelecky:2006ta} & polarization measurement in gamma ray bursts\\\hline
		$10^{-9}$ &$2007$& \cite{Muller:2007es} & {atomic gravimeter}\\\hline
		$10^{-15}$ &$2004$& \cite{Wolf:2004gg} &comparison of a cryogenic sapphire microwave resonator and a hydrogen maser\\\hline
		$10^{-18}$ & $2014$&\cite{Nagel:2014aga} &terrestrial Michelson-Morley experiment\\\hline
		$10^{-21}$ &$2018$& \cite{Sanner:2018atx} & Michelson-Morley with trapped ions {(assuming no Lorentz-symmetry violation for electrons)}\\\hline
		$10^{-20}$ &$2016$& \cite{Kostelecky:2016kkn} & light interferometry (LIGO data)\\\hline
	\end{tabular}
	\caption{Different experimental bounds on the analogue of our Lorentz-symmetry breaking coupling $\zeta$ for the photon sector of the Standard Model.  We assume that the existence of a single vector field $n_{\mu}$ as source of a preferred frame is the only source of Lorentz-symmetry violations. In this case the coupling $\zeta$ is the unique non-zero coupling. For each experiment, the strongest bound on the coefficients of $k_{\rm F}^{\mu\nu\rho\sigma}$, cf.~\eqref{eq: gaugeLIV}, are translated into bounds on $\zeta$. Except for the second line, all bounds assume the absence of LIV couplings in the pure gravity sector, since some assumption on the gravitational background is necessary for the conversion of experimental data to bounds on LIV couplings. We stress the difference between the experimental bounds on the photon-LIV coupling and the coupling $\zeta$ in our toy model. The above experimental bounds on the photon LIV couplings are intended to give an impression on the sensitivity of experiments in the photon sector. They do not directly translate into constraints on the LIV coupling $\zeta$ in our toy model.}
	\label{tab: LIVMatterConstraints}
\end{table*}

As for the matter part of the action, we focus on the Abelian gauge sector with
\begin{equation}
\Gamma_{k}^{\rm Abelian}
=\frac{Z_A(k)}{4}\int \sqrt{\det(g_{\kappa\epsilon})} \,\,g^{\mu\rho}g^{\nu\sigma}F_{\mu\nu}F_{\rho\sigma},
\label{eq:gaugekinetic}
\end{equation}
where $F_{\mu\nu}$ is the field-strength tensor of the Abelian gauge field $A_{\mu}$ and $Z_A(k)$ is a wave-function renormalization of the gauge field. Even in the absence of charged matter, quantum fluctuations of gravity generate a non-trivial scale-dependence, giving rise to an anomalous dimension
\be
\eta_A = - k\, \partial_k \ln Z_A(k).
\ee
Finally, all possible extensions of the Abelian gauge sector that violate Lorentz invariance but preserve CPT and gauge symmetry can be written as \cite{Colladay:1998fq}
\begin{equation}
\Gamma_{k}^{\rm Abelian,\, LIV}
=\frac{Z_A(k)}{4}\int \sqrt{\det(g_{\kappa\epsilon})}\,\, k_{\rm F}^{\mu\nu\rho\sigma}(k)\,F_{\mu\nu}F_{\rho\sigma},
\label{eq: gaugeLIV}
\end{equation}
where $k_{\rm F}^{\mu\nu\rho\sigma}(k)$ is real and has the symmetries of the Riemann tensor, i.e., antisymmetry under $\mu \leftrightarrow \nu$ and $\rho \leftrightarrow \sigma$ and symmetry under an exchange of the pairs~$\{\mu,\nu\}$ and~$\{\rho, \sigma\}$. To see this, start with a general tensor $\hat{k}_{\mu\nu\rho\sigma}$ with no symmetries. Its completely antisymmetric part results in the CP-violating~$\tilde{F}F$-term, which is a total derivative for the Abelian gauge field. Further, symmetry under exchange of the index pairs~$(\mu, \nu) \leftrightarrow (\rho,\sigma)$ follows from the contraction with two field-strength tensors. Finally, gauge symmetry demands antisymmetry of the field-strength tensor, resulting in antisymmetry of~$k_{\mu\nu\rho\sigma}$ under exchanges of indices~$\mu \leftrightarrow \nu$.

The presence of the LIV operator in Eq.~\eqref{eq: gaugeLIV} leads to vacuum birefringence: The dispersion relation resulting from Eqs.~\eqref{eq:gaugekinetic} and \eqref{eq: gaugeLIV} is still linear in the spatial momentum, i.e., $p_0 =c(k_F)\, |\vec{p}|$. Therefore, there is no wavelength dependence in the speed of propagation. Yet, the two polarizations feature a different proportionality factor $c(k_F)$, which leads to a phase shift between the two polarizations that accumulates with propagation distance. For a detailed discussion, see
 \cite{Kostelecky:2001mb, Kostelecky:2002hh}.
 
Under the impact of quantum fluctuations, $k_{\rm F}^{\mu\nu\rho\sigma}$ acquires a dependence on the RG scale $k$. For a general dynamical preferred frame \cite{Jacobson:2000xp}, which was explicitly applied to  Horava-gravity \cite{Horava:2009uw,Contillo:2013fua,DOdorico:2014tyh,DOdorico:2015pil,Barvinsky:2015kil,Barvinsky:2017kob,Barvinsky:2019rwn} in \cite{Bluhm:2019ato}, the only possible tensor in the general expression Eq.~\eqref{eq: gaugeLIV} is\footnote{The couplings associated with the breaking of Lorentz symmetry in \cite{Bluhm:2019ato} and in our work are related via $\frac{1}{2}(1-\lambda_{\gamma})=\zeta$.}
\begin{equation}
k_{\rm F}^{\mu\nu\rho\sigma}=\frac{\zeta}{4} \Bigl(n^{\mu}n^{\rho}g^{\nu\sigma}+n^{\nu}n^{\sigma}g^{\mu\rho} - n^{\nu}n^{\rho}g^{\mu\sigma} - n^{\mu}n^{\sigma}g^{\nu \rho}\Bigr),
\label{eq: gaugeLIVV}
\end{equation}
with the coupling $\zeta= \zeta(k)$. We stress that in the presence of a single vector field $n_{\mu}$ as the source of a preferred frame, $\zeta$ is the unique coupling that can be nonzero. In particular, if we assumed that the different components of~$k_F$ were parameterized by different couplings, this would amount to the introduction of the corresponding nontrivial \emph{tensor} as the source of a preferred frame\footnote{Experimentally, the various components of $k_F$ are constrained individually, as typically no assumption on the precise way in which a preferred frame is selected, is made.}.
Therefore, all experiments that put constraints on individual components of the general tensor $k_{\rm F}^{\mu\nu\rho\sigma}$ in Eq.~\eqref{eq: gaugeLIV} automatically put constraints on the coupling $\zeta$. 
We summarize experimental bounds in Table~\ref{tab: LIVMatterConstraints}, where the strongest bound on any of the components of $k_{\rm F}^{\mu\nu\rho\sigma}$ is translated into a bound on~$\zeta$. Note that although the measurements on LIV in the matter sector could be affected by the presence of higher-order operators, due to the Planck-scale suppression of the corresponding couplings their impact to the low-energy measurement of $k_F$ is negligible.

\section{The relation between Lorentz invariance violation in the gravity sector and in the matter sector}
\label{sec: GravMatInterplay}

It is a crucial question, whether Lorentz symmetry breaking necessarily percolates from the gravitational sector into the matter sector. This matters both for theoretical as well as phenomenological reasons: On the theoretical side, this is crucial to understand the form of a matter sector that is consistently coupled to a LIV-gravity sector. On the phenomenological side this is key, as it allows to translate strong observational bounds on LIV in the matter sector into constraints on LIV couplings in the quantum gravity sector.

More formally, the key question  is whether the Lorentz invariant subspace of the matter sector is attractive or repulsive under the RG flow towards low energies. In other words, starting from small deviations from the Lorentz invariant hypersurface at large energies, is the system driven away from the symmetric subspace or towards it, when lowering the energy?
To answer this question as comprehensively as possible, we remain agnostic about the properties of a UV completion for the system. Thus, we view the description in terms of quantum field theory as an effective description with a high-energy (i.e., transplanckian) cutoff scale $k_{\rm i}\gg M_{Pl}^2$. At that cutoff scale, a microscopic model sets the initial conditions for the RG flow towards the IR by determining the values of couplings at that scale\footnote{In case of an asymptotically safe/free fixed point, that initial condition corresponds to values of the relevant couplings a high-energy scale at which relevant perturbations drive the system away from the fixed point.}. We explore, whether gravitational fluctuations then drive the LIV matter coupling back to zero, or whether there is a nonzero preferred value.

\begin{figure}[!t]
\centering
	\includegraphics[width=\linewidth]{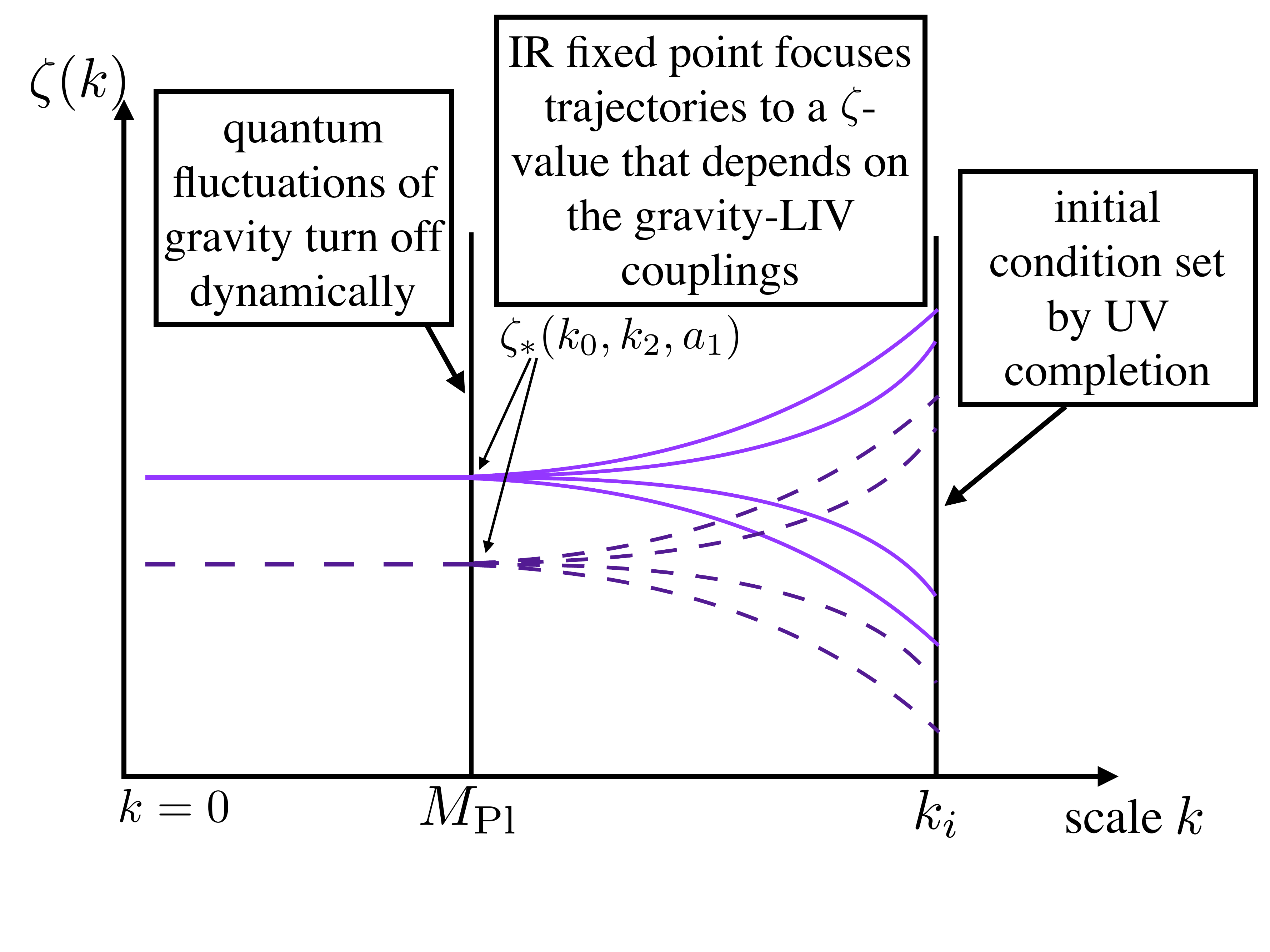}
	\caption{\label{fig:schematic}We show a schematic illustration of the key idea underlying our results.}
\end{figure}

In a nutshell, our results are the following: We will show that $\zeta(k)$ cannot consistently be set to zero in the presence of $k_0$, $k_2$ and $a_1$ at high energies. This is a consequence of the absence of a free fixed point in the beta function for $\zeta$. In other words, quantum fluctuations generate $\zeta$ and drive it away from zero. Moreover we find that, within our approximation, there is always an IR-attractive fixed point at a finite value of~$\zeta$. Consequently, quantum fluctuations drive $\zeta$ towards a preferred, nonzero value. Under the RG flow, a large range of initial conditions~$\zeta(k_{\rm i})$, set at the ultraviolet scale~$k_{\rm i}$, 
is thus mapped into a unique Planck-scale value, corresponding to the IR-attractive fixed point of the RG flow. 
Below the Planck scale, the effect of $k_0$, $k_2$ and $a_1$ switches off dynamically, simply because quantum fluctuations of gravity become negligible and quantum gravity decouples from particle physics.
In our toy-model, the flow of $\zeta$ vanishes in this regime, as quantum fluctuations of gravity are the only ones that drive the system. Therefore, the universal fixed-point value attained by the RG flow at Planckian scales is also the low-energy value of $\zeta(k=0)$, cf.~Fig.~\ref{fig:schematic} for an illustration. 

In a more complete treatment that accounts for the other Standard-Model degrees of freedom, additional fluctuations would drive the low-energy flow of~$\zeta$. Since $\zeta$ is a marginal coupling, below the Planck scale it is expected to depend logarithmically on the RG scale $k$. In our toy model, this low-energy running is absent. 

The low-energy value of $\zeta$ (referring to the actual electromagnetic interaction of the Standard Model) is constrained observationally. In turn, this experimental bound can be mapped onto a constraint for its Planck-scale value. We expect that the latter is a function of the LIV-gravity couplings $k_0$, $k_2$ and $a_1$, just as it is in our toy model. Accordingly, observational constraints on $\zeta$ constrain the microscopic values of $k_0$, $k_2$ and $a_1$, and can therefore indirectly constrain the fundamental symmetries of the theory. The conditions under which such an indirect constraint arises will be discussed in detail below.

This idea constitutes an example of how studies of the interplay of quantum gravity with matter can be key to constrain quantum gravity observationally, by tapping into the wealth of experimental data on particle physics.

Having explained the key idea underlying our work, we can now discuss in detail how the IR-attractive fixed point is generated and how the gravitational couplings impact the scale dependence of the LIV-matter coupling~$\zeta$. To this end we compute the scale dependence of~$\zeta$, treating all gravitational couplings as input parameters. (See \cite{Knorr:2018fdu} for the beta functions of these couplings.)
This allows us to remain agnostic about a UV completion for the gravity sector\footnote{In \cite{Kostelecky:2003fs} the spontaneous breaking of Lorentz invariance has been discussed, while recent studies of gravity-matter systems also allow for the possibility of explicit symmetry breaking, cf.~\cite{Bluhm:2019ato,Bluhm:2019hct}.}, and explore a large class of possible models, labeled by different values of these couplings, simultaneously.
To obtain an analytical expression for the scale dependence of~$\zeta$, we expand all expression to first order in $k_0$, $k_2$ and $a_1$ and to second order in~$\zeta$, which is sufficient for the assumption of small deviations from the Lorentz invariant hypersurface at high energies. 
In fact, it is the dimensionless ratios\footnote{Note that here $g(k)$ stands for the dimensionless Newton coupling, and should not be confused with the ``g'' of the metric tensor,~$g_{\mu\nu}$, which we always refer to with two indices.}
\be
g(k)= G_N(k)\, k^2, \quad \lambda(k) = \Lambda(k)\, k^{-2},
\ee
which enter the beta function
\be
\beta_{\zeta} = k\,\partial_k\, \zeta(k).
\ee
For technical details of our calculation, we refer the reader to the appendix. To evaluate the RG flow, we use the Mathematica package \emph{xAct} \cite{Brizuela:2008ra,Martin-Garcia:2007bqa,MartinGarcia:2008qz,2008CoPhC.179..597M} as well as the FORM-tracer~\cite{Cyrol:2016zqb}.

Driven by quantum fluctuations of gravity and the Abelian gauge field itself, the beta function for $\zeta$ is
\bea
\beta_\zeta&=&g\,\bigg(-\frac{10 a_1+21 k_0+257 k_2}{384 \pi  (1-2 \lambda )^2}+\frac{-6 a_1+53 k_0+329 k_2}{576 \pi  (1-2 \lambda )^3}\bigg)\notag\\
&{}&+\zeta\,  g\, \bigg(\frac{1}{6 \pi  (1-2 \lambda )}\notag\\
&{}&\hspace{25pt}-\frac{183 a_1-390 k_0-1690 k_2+1840}{960 \pi  (1-2 \lambda )^2}\notag\\
&{}&\hspace{25pt}+\frac{2313 a_1-5 (246 k_0+4039 k_2)}{1440 \pi  (1-2 \lambda)^3}\bigg)\notag\\
&{}&+\zeta ^2\, g\,\bigg(
\frac{79}{60 \pi  (1-2 \lambda )}\notag\\
&{}&\hspace{27pt}-\frac{21 a_1+495 k_0-920 k_2+5072}{960 \pi  (1-2 \lambda )^2}\notag\\
&{}&\hspace{27pt}+\frac{6911 a_1-9515 k_0-60420 k_2}{1440 \pi  (1-2 \lambda )^3}\bigg),
\label{eq: betazeta}
\eea
where we have dropped the dependence on~$k$ from all couplings for brevity. We briefly highlight the existence of the nontrivial denominators which are in contrast to beta functions obtained with perturbative techniques, which are typically purely polynomial. Such denominators arise when there is a mass-like term for a field, and result in a dynamical decoupling of the corresponding degree of freedom once the RG scale~$k$ drops below the corresponding mass. For metric fluctuations, the cosmological constant acts akin to a mass-like term, suppressing metric fluctuations for large negative~$\lambda$. Notice that this refers to the microscopic (i.e., high-energy) value of the dimensionless cosmological constant, which is itself a  scale-dependent coupling that can take a rather different value in the UV than in the IR. In particular, a negative $\lambda$ in the UV is not incompatible with a positive cosmological constant at observational scales~\cite{Dona:2013qba,Biemans:2017zca}.

In order to understand the implications of the expression~\eqref{eq: betazeta}, we focus on special cases first. 

If we set $g=0$, then the entire beta function vanishes. This is a consequence of the fact that at $g=0$, the model consists of just a kinetic term for the Abelian gauge field, i.e., it is a noninteracting theory. The beta functions in such a theory vanish identically. Beyond our toy model, the existence of additional matter degrees of freedom would not change this conclusion, unless there was LIV already present in other couplings in the matter sector. 

At $g\neq 0$, we focus on the limit $\zeta = 0$ first. In this case, only the first line in Eq.~\eqref{eq: betazeta} remains. Except for very special points in the parameter space spanned by~$\{a_1, k_0, k_2, \lambda\}$, this expression is nonvanishing. This has important implications: Even setting $\zeta(k_{i})=0$ (where $k_{\rm i}\gg M_{Pl}$ is an arbitrary initial scale),~$\beta_{\zeta}(\zeta=0)\neq 0$, and therefore $\zeta(k_{\rm i}-\delta k)\neq 0$. In other words,  quantum fluctuations generate $\zeta$, even if  it vanishes at $k_{\rm i}$. 
On the other hand, if $k_0=0, k_2=0, a_1=0$, then the first line of Eq.~\eqref{eq: betazeta} vanishes identically. This choice corresponds to a gravitational theory that respects full diffeomorphism invariance. In this case, there is no Lorentz symmetry violation in the gravitational sector, which is reflected in the existence of a fixed point of $\beta_{\zeta}$ at $\zeta=0$. The hypersurface in the space of couplings that preserves Lorentz symmetry in the matter sector is only an invariant surface under the RG flow, if no LIV exists in the gravity sector. Hence, Lorentz-symmetry breaking will percolate from the gravity sector into the matter sector, if the couplings $k_0,k_2$ and $a_1$ are non-vanishing.

In the next step, we take the terms $\sim \zeta\, g$ and~$\sim \zeta^2\, g$ in the second and following lines of Eq.~\eqref{eq: betazeta} into account. The beta function $\beta_{\zeta}$ is generically nonzero, i.e., starting from an initial condition $\zeta(k_{\rm i})$, the LIV coupling $\zeta(k)$ will flow, and  assume a different value at lower scales. In this context, the notion of attractors of the flow, i.e., fixed points, is crucial. Under the influence of an IR-attractor, a large interval of initial conditions in the UV is mapped to a small interval of values at lower scales: A \textit{universal} prediction of a nonzero value of $\zeta$ arises that is largely independent of the UV initial conditions. We now analyze the notion of such IR attractors in terms of fixed points and pseudo fixed points in more detail.

\begin{figure}
\centering
	\includegraphics[width=\linewidth]{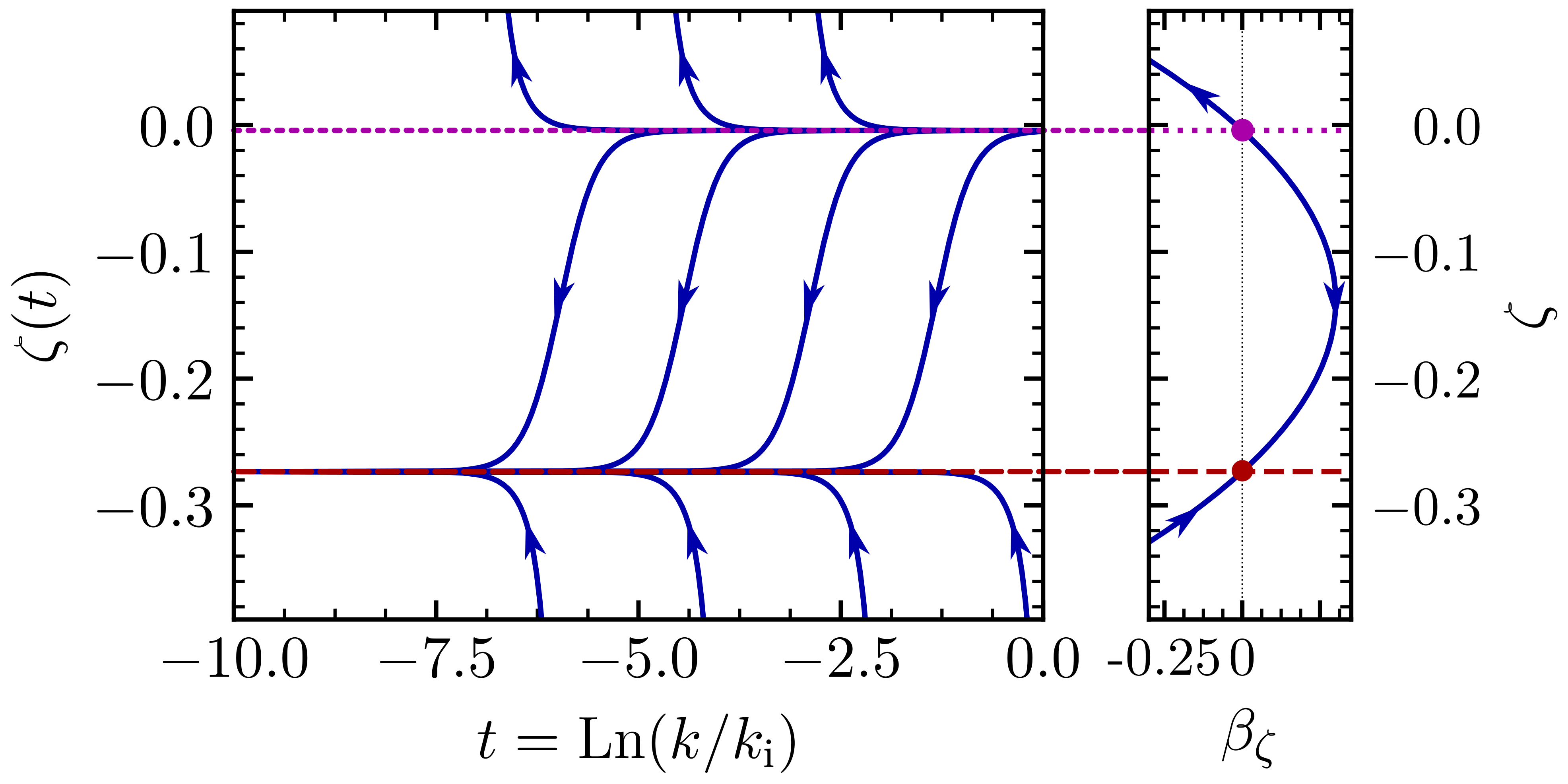}
	\caption{\label{fig:fancyplotbeta} We show the beta function (right panel) and the associated flow of $\zeta(k)$ for $g=1$, $a_1=\lambda=0$ and $k_2=k_0=1$ (left panel). The magenta dotted line corresponds to the IR-repulsive fixed point. The blue lines are a sample of RG trajectories obtained from varying the initial condition~$\zeta(k_{\rm i})$. The arrows indicate the direction of the flow, towards the IR. RG trajectories with initial conditions above the magenta line are driven away from the Lorentz-invariant sub-theory space. Conversely, RG trajectories set by initial conditions below the magenta line flow towards the IR-attractive fixed point, i.e., the red dashed line, at low energies.}
\end{figure}

Let us schematically write
\be
\label{eq: betaschem}
\beta_{\zeta} = b_0+ b_1\, \zeta + b_2\, \zeta^2,
\ee
and let us treat the $b_i$ as (real) constants for now.
The zeros of $\beta_{\zeta}$, where the scale-dependence of $\zeta$ vanishes, are
\be
\label{eq: FPsec3}
\zeta_{*, 1/2}= \frac{-b_1\pm\sqrt{b_1^2-4b_0 b_2}}{2b_2}.
\ee
If $b_1^2-4b_0\, b_2>0$, these fixed points of the RG flow lie at real values which are generically nonzero. One of the two fixed points is IR-repulsive and the other is IR-attractive, as one can infer by calculating the critical exponents
\be \label{eq: critexp}
\theta_{1/2}= -\frac{\partial \beta_{\zeta}}{\partial \zeta}\Big|_{\zeta_{\ast, 1/2}}=\mp \sqrt{b_1^2-4b_0\,b_2}.
\ee
The critical exponent encodes whether a fixed point is IR attractive or IR repulsive.
A positive critical exponent signifies that the distance to the corresponding fixed point increases under the RG flow to the IR -- the coupling is a relevant perturbation of this fixed point. In contrast, a negative critical exponent implies that the fixed point is an attractor of the RG flow. 
This is clearly visible in Fig.~\ref{fig:fancyplotbeta} where we show selected RG trajectories (blue lines) for~$\zeta(k)$. The fixed point coming with $\theta_i<0$ (red dashed line) acts as an attractor, whereas the fixed point with~$\theta_i>0$ (magenta dotted line) repulses RG trajectories. Therefore, a \textit{universal} prediction arises: The IR-attractive fixed point at~$\zeta_{\ast, 1}$ \emph{focuses} trajectories. 
Except for initial conditions \footnote{{This follows as an IR repulsive fixed-point shields a certain set of UV initial conditions from the IR attractive fixed point at lower energies. Specifically, in the case of Fig.~\ref{fig:fancyplotbeta}, any value of~$\zeta(k)<\zeta_{\ast, 2}$, with~$\zeta_{\ast, 2}$ being the IR repulsive fixed point, is inaccessible from initial conditions~$\zeta(k_{\rm i})>\zeta_{\ast, 2}$.}} which lie above~$\zeta_{\ast,2}$, a large range of initial conditions at $k=k_{\rm i}$ is mapped to $\zeta (10^{-10} k_{\rm i})=\zeta_{\ast, 1}$.
Therefore, trajectories with initial conditions below the IR-repulsive fixed point (magenta dotted line) will be focused on the IR-attractive fixed point $\zeta_{\ast, 1}$. 
Trajectories starting at $\zeta(k_{\rm i})>\zeta_{\ast, 2}$ are  quickly driven towards rather large values of $\zeta$.

\begin{figure}
\centering
	\includegraphics[width=\linewidth]{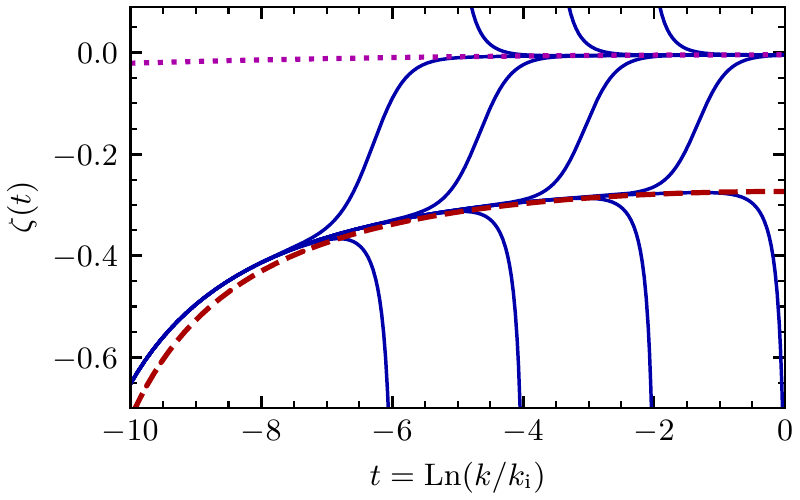}
	\caption{\label{fig:flowplotrunning} We show the flow for the case $g=1$, $\lambda=0$, $a_1=0$, $k_0=1-t^2/20$ and $k_2=1$. The zeros of $\beta_{\zeta}$ acquire a scale dependence through the scale-dependence of the LIV coupling~$k_0$. These pseudo-fixed points (magenta dotted line and red dashed line) approximate the attractors/repulsors of the flow. 
	}
\end{figure}

Next, we have to understand the situation when the coefficients~$b_0$,~$b_1$ and~$b_2$ are scale dependent due to the scale dependence of the gravitational couplings. In this case, $\zeta_{\ast,1/2} = \zeta_{\ast,1/2}(k)$ become pseudo fixed-points: They are still the solutions to $\beta_{\zeta}=0$, but these solutions are no longer scale-independent. Accordingly, they lose the interpretation as a scale-invariant regime of the theory, but they do keep the interpretation as (scale-dependent) attractors and repulsors of the flow\footnote{{Strictly speaking, not the points $\zeta_{\ast,1/2}(k)$ are the attractors/repulsors, but a close-by set of points $\tilde{\zeta}_{\ast,1/2}(k)$, where the slope of $\tilde{\zeta}_{\ast,1/2}(k)$ balances a non-vanishing contribution from $\beta_{\zeta}$. The larger the slope of the beta function compared to the ``speed" of the points $\tilde{\zeta}_{\ast,1/2}(k)$, the closer~$\tilde{\zeta}_{\ast,1/2}(k)$ lie to the pseudo fixed points $\zeta_{\ast,1/2}(k)$. }}. Their effectiveness depends on the speed of the flow -- the derivative of the beta function -- compared to the speed with which the pseudo-fixed-point value changes as a function of the scale. 
If the derivative of the beta function is large, then the flow easily follows the IR-attractive pseudo fixed-point, as in the example  in~Fig.~\ref{fig:flowplotrunning}. From an appropriate set of initial conditions, the flow is quickly attracted to the vicinity of the IR-attractive pseudo fixed-point, and then converges to it at lower scales. In the other case, where the speed of the flow is slow compared to the rate of change of the pseudo-fixed-point value, the flow cannot follow the pseudo fixed-point and the latter becomes ineffective as an attractor of the flow. 

In the following, we will make the assumption that the gravitational couplings change as a function of the RG scale in such a way that the rate of change of the pseudo fixed-point is smaller than the speed of the flow. This ensures that the pseudo fixed-points are effective as attractors/repulsors. For an IR-attractive pseudo fixed point, a large range of UV initial conditions for~$\zeta(k)$ is mapped to a small interval around the pseudo fixed-point (red dashed curve, Fig.~\ref{fig:flowplotrunning}). The latter is the \emph{instantaneous} fixed-point value, i.e., the solution to~$\beta_{\zeta}=0$ with the values of the gravitational couplings at the Planck scale. Therefore, if the initial condition $\zeta(k_{\rm i})$ lies in the basin of attraction of the IR-attractive pseudo fixed-point, the ``history" of the trajectory, i.e., the scale-dependence of the gravitational couplings above the Planck scale, becomes unimportant. Otherwise, i.e., if $\zeta(k_{\rm i})$ is outside the basin of attraction of the IR-attractive pseudo fixed-point, the corresponding RG trajectory will flow away from the IR-repulsive fixed point (magenta dotted curve, Fig.~\ref{fig:flowplotrunning}), so that $\zeta(k)$  becomes large at low energies.

As the gravitational contributions to the flow turn off at the Planck scale, the flow towards the IR vanishes. Therefore, each value at the Planck scale can be translated into a unique value in the IR, such that the IR value of $\zeta$ is a prediction of the theory. For initial conditions in the basin of attraction of the IR-attractive pseudo fixed point, the IR value of $\zeta$ is a universal\footnote{Universality here does not refer to scheme-independence, as the gravitational contributions depend on the scheme due to the dimensionful nature of the Newton coupling. Universality in our context means the independence from microscopic physics, i.e., initial conditions for the RG flow.} prediction, as in this case the flow ``looses memory'' of the initial conditions: The IR-attractive pseudo fixed point  depends on the gravity-LIV couplings of the system, but is independent of the initial value $\zeta(k_{\rm i})$. Thus, changes in the gravity LIV couplings result in a change of the Planck-scale value of $\zeta$, and thereby its low- scale value.\\

In summary, this setup provides us with a map between Planck-scale values of the gravitational couplings and the IR value of $\zeta$. With the help of such a map, strong observational constraints on $\zeta$ can in principle be translated into strong constraints on the gravitational couplings. 
These hold in a setting where 
\begin{enumerate}
	\item a quantum-field theoretic description is applicable beyond the Planck scale, 
	\item the rate of change of the pseudo fixed point is smaller than the speed of the flow (this can be checked from the beta function, given a particular scale-dependence for the gravitational couplings), 
	\item  the initial condition for $\zeta(k_{\rm i})$ lies in the basin of attraction of the IR fixed point,
	\item the additional Standard Model degrees of freedom beyond our setting do not significantly alter the flow of $\zeta$.
\end{enumerate}
As already mentioned, for initial conditions outside the basin of attraction of the IR fixed point, the RG flow generically drives $\zeta$ towards large absolute values. In this case, the IR value of $\zeta$ is not a universal prediction, but depends on the initial condition $\zeta(k_{\rm i})$. Generically, the initial condition~$\zeta(k_{\rm i})=0$ results in 
\be
\zeta(k) = -\frac{b_0}{b_1} \left(1-\left(\frac{k_{\rm i}}{k}\right)^{b_1} \right),
\ee
which holds for small enough $\zeta$, so that the quadratic term in Eq.~\eqref{eq: betaschem} can be neglected. Parametrically, the value of $\zeta(k)$ is set by the gravitational LIV couplings which enter~$b_0$ and~$b_1$.
To keep $\zeta$ small, it follows that $b_0\ll1$ if $b_1 \sim \mathcal{O}(1)$, or $b_1\ll1$. Such conditions require either very small LIV gravity couplings, and for the case of $b_1 \sim \mathcal{O}(1)$ at least one LIV coupling of order 1, which is incompatible with direct constraints \cite{Gumrukcuoglu:2017ijh}, derived under the assumption of photons propagating at the speed of light, see \cite{Jacobson:2001tu,Jacobson:2002ye,Bolmont:2006kk,Ackermann:2009aa,Vasileiou:2013vra,Kostelecky:2013rv,Ellis:2018lca,Abdalla:2019krx}. 
Accordingly, we tentatively conclude that strong observational constraints on $\zeta$ are only compatible with Lorentz invariance violation~$\mathcal{O}(1)$ in the gravitational sector under the assumption of very special initial conditions for~$\zeta(k_{\rm i})$. 

\section{Constraints}
\label{sec: constraints}
We now discuss how the considerations in the previous section constrain the quantum gravitational LIV sector. To exemplify this, we work with the beta function in Eq.~\eqref{eq: betazeta}. This beta function was obtained within truncations of the dynamics  and further approximations, cf.~App.~\ref{sec: AppFRG}, and is therefore subject to systematic errors.
Further, our results are obtained using a Euclidean setup and we make the assumption that the form of the beta function will be the same in the Lorentzian setup. Due to the absence of a well-defined Wick rotation in quantum gravity, this is an assumption which is not straightforward to test, although the presence of a foliation is a prerequisite for a Wick rotation. Finally, we work within a toy model for the Standard Model which only includes an Abelian gauge sector, but neglects the other Standard Model degrees of freedom as well as the difference between the Abelian hypercharge gauge field and the photon that is due to electroweak symmetry breaking.
Due to the above points, the quantitative limitations of our study should be obvious. Nevertheless,  the result  that the gravity LIV couplings enter the beta function for matter LIV couplings with numerical prefactors $\mathcal{O}(1)$ should be generic. Therefore, the key result, that constraints on the matter LIV couplings of order $\mathcal{O}(10^{\#})$ constrain gravitational LIV couplings to roughly the same precision, is expected to be generic and is indeed a key point we want to make in this paper.

\begin{figure*}[t!]
	\includegraphics[width=\linewidth]{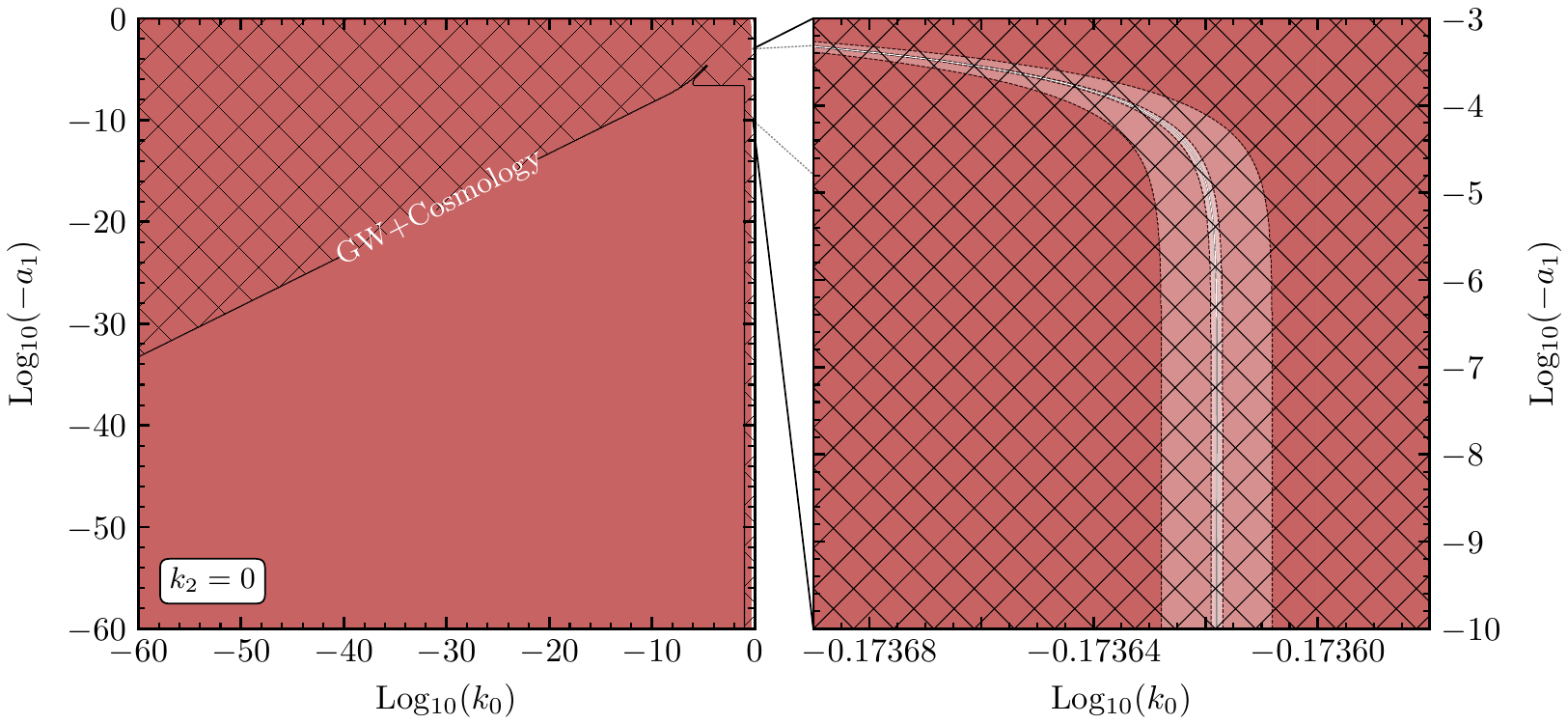}
	\caption{Exclusion for $\lambda>\lambda_{\rm crit}$ with initial conditions in the basin of attraction of the IR attractive interacting fixed point $\zeta_{*,2}$, for $\lambda=k_2=0$. Left panel: the red region shows the excluded region by demanding that $\zeta_{*,2}<10^{-4}$. The hatched area marks the region which is already excluded by cosmology and observation of gravitational waves. Right panel: zoom into the only region, which can accommodate $\zeta_{*,2}<10^{-4}$ (lighter red areas). The tiny white band corresponds to values of~$k_0$ and~$a_1$ that make~$\zeta_{*,2}$ exactly zero, according to Eq.~\eqref{eq: IntFP}. This region is already excluded by cosmological observations \cite{Gumrukcuoglu:2017ijh}.}
	\label{fig: ExclPlotNGFPk20}
\end{figure*}

Since we are interested in small deviations from the Lorentz-invariant subspace, let us start with the Lorentz invariant case, i.e., $k_0=k_2=a_1=0$. In this case the coefficient $b_0$ in Eq.~\eqref{eq: betaschem} vanishes, and $\beta_{\zeta}$ features one Gaussian and one non-Gaussian fixed-point:
\begin{equation}
\label{eq: newFP}
(\zeta_{*,\,1},\zeta_{*,\,2})\big|_{k_0=k_2=a_1=0}=\left(0, -\frac{5 (4 \lambda +21)}{158 \lambda +238}\right)\,,
\end{equation}  
with critical exponent
\begin{equation}
\label{eq: critexpsym}
(\theta_1,\theta_2)\big|_{k_0=k_2=a_1=0}=\left(\frac{g (4 \lambda +21)}{12 \pi  (1-2 \lambda )^2},-\frac{g (4 \lambda +21)}{12 \pi  (1-2 \lambda )^2}\right)\,.
\end{equation} 
For non-vanishing, but small LIV couplings $k_0,\,k_2$ and $a_1$, the coefficient $b_0$ in Eq.~\eqref{eq: betaschem} is non-vanishing, shifting the Gaussian fixed point (GFP)  $\zeta_{*,\,1}$ to an interacting shifted Gaussian fixed point (sGFP). (The notation $\zeta_{*,\, 1/2}$ in Eq.~\eqref{eq: newFP} was chosen such that $\zeta_{*,\,1}$ always corresponds to the GFP, in contrast to Eq.~\eqref{eq: FPsec3}, where the sGFP can be either of the fixed points, depending on the sign of $b_1$.) For small LIV gravity couplings, the sGFP is a continuous deformation of the GFP in the symmetry-restored case, while the fixed point $\zeta_{*,\,2}$ is always interacting. Therefore, for small LIV gravity couplings, the existence of the sGFP is robust and controlled as it is a continuous deformation of the free fixed point. The interacting fixed point $\zeta_{*,\,2}$, however, cannot be traced back to a free fixed point. Therefore, its existence might be subject to extensions of the truncation.\\
From Eq.~\eqref{eq: critexpsym} it is evident that the critical exponent of the (s)GFP changes sign at $\lambda_{\rm crit}=-\tfrac{21}{4}$. For $\lambda>\lambda_{\rm crit}$  the sGFP is IR repulsive and the interacting fixed point IR attractive. This situation is illustrated in Figs.~\ref{fig:fancyplotbeta} and \ref{fig:flowplotrunning}.

In the following we investigate both cases, i.e.,~$\lambda>\lambda_{\rm crit}$ and~$\lambda<\lambda_{\rm crit}$.\\

\paragraph{Constraints on LIV gravity couplings for $\lambda>\lambda_{\rm crit}$}\textcolor{white}{b}\\
In the case of $\lambda>\lambda_{\rm crit}$, the sGFP $\zeta_{*,1}$ is IR repulsive, while the interacting fixed point $\zeta_{*,2}$ is IR attractive, as illustrated in Fig.~\ref{fig:fancyplotbeta}. The basin of attraction of~ $\zeta_{*,2}$ contains all values~$\zeta(k)<\zeta_{*,1}$. For all these values it follows that $\zeta(M_{\rm Pl})\approx\zeta_{*,2}$. 
To link to experimental constraints,~$\zeta(M_{\rm Pl})$ has to be used as the initial condition for the RG flow below the Planck scale. In our case, the dynamical switching off of gravitational fluctuations, encoded in a quadratic scaling of~$g(k)$ to zero, results in a swift freeze-out of~$\zeta(k<M_{\rm Pl})$.
Therefore, for the exclusion plots in Fig.~\ref{fig: ExclPlotNGFPk20} and Figs.~\ref{fig: ExclPlotPosNGFP}--\ref{fig: k2posexclreg}, we  make the rather conservative assumption that the ratio~$\zeta(0)/\zeta(M_{\rm Pl})$ is not smaller than~$1/10$.

Through the map $\{k_0, k_2, a_1, \lambda\} \rightarrow \zeta_{\ast , 2} = \zeta(M_{\rm Pl}) \rightarrow \zeta(k=0)$, we translate the experimental constraints on the LIV coupling $\zeta$ to constraints on $\zeta$ at the Planck scale. In turn, this constrains the gravitational LIV couplings~$k_2$, $k_0$ and $a_1$. We emphasize that the limitations of our study should be kept in mind when interpreting the constraints that result from our study.

To highlight the typical strength of such constraints, let us investigate the special case of $\lambda=0$. To linear order in the LIV couplings, the IR-attractive fixed point is given by
\begin{equation}
\label{eq: IntFP}
\zeta_{*,2}=\frac{-179025 a_1+766741 k_0+1776274 k_2}{1165248}-\frac{15}{34}\,.
\end{equation}
If we assume experimental constraints $|\zeta| < 10^{-10}$, it is evident from Eq.~\eqref{eq: IntFP} that values for~$k_2,\, k_0$ and~$a_1$ of order~1 are already excluded.
Only specific combinations of $k_0,\,k_2$ and $a_1$ with at least one coupling of $\mathcal{O}(1)$ can satisfy this bound, cf.~Fig.~\ref{fig: ExclPlotNGFPk20}. 
This is a direct consequence of the non-vanishing value of~$\zeta_{*,2}\big|_{k_0=k_2=a_1=0}$, which is of~$\mathcal{O}(10^{-1})$.  However, as shown in~Fig.~\ref{fig: ExclPlotNGFPk20}, any~$\mathcal{O}(1)$ LIV gravity coupling compatible with~$|\zeta|<10^{-10}$ is already excluded by direct cosmological constraints and by the observational data on gravitational waves \cite{Gumrukcuoglu:2017ijh}. We emphasize that the existence of the IR attractive fixed point $\zeta_{*,2}$ saves the system from an uncontrolled behavior towards the IR, rather than generating the strong constraints. In other words, without the interacting fixed point, the system would be driven to even larger values of couplings in the IR, resulting in even stronger constraints.
If therefore in future studies the interacting fixed point $\zeta_{*,2}$ turns out to be spurious tiny violations of Lorentz invariance in the gravity sector will conflict with constraints in the matter sector, since $\zeta$ will increase very strongly towards the IR.

\begin{figure}[t!]
	\centering
	\includegraphics[width=\linewidth]{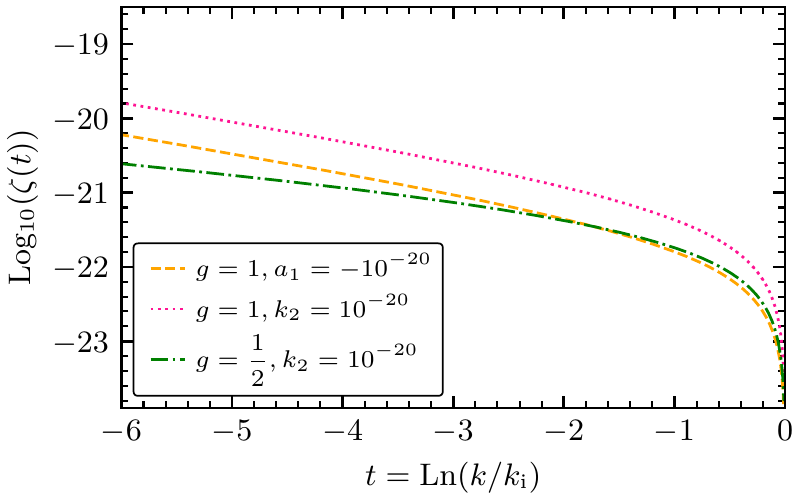}
	\caption{For the case $\lambda=0$ (which is generic for the purposes of this plot), the RG flow over a few orders of magnitude generates~$\zeta(k)\sim c$ from $\zeta(k_{\rm i})=0$, for any non-vanishing gravity LIV coupling of~$\mathcal{O}(c)$. The couplings $k_0$, $k_2$ and $a_1$ not mentioned in the respective label are set to zero.}
	\label{fig: RegenFLow}
\end{figure}

For initial conditions in the range~$\zeta(k_{\rm i})\geq \zeta_{*,1}$, the flow of~$\zeta(k)$ is governed by the sGFP $\zeta_{*,1}$, which is IR-repulsive.  In contrast to the IR-attractive fixed point, the sGFP defocuses trajectories. Hence, $\zeta(k)$ is driven away from $\zeta_{*,1}$ towards lower scales, and no universal bound arises. In this case, the IR value of $\zeta$ is generically too large to stay within the experimental bounds. The set of initial conditions at the scale~$k_{\rm i}$ which allow for a small enough $\zeta (M_{\rm Pl})$ to satisfy strong constraints is very special: Generically, the flow towards larger $|\zeta|$ is rather fast, unless one starts very close to the sGFP. Specifically, for a non-vanishing value $c$ of any of the gravity-LIV couplings, a value $\zeta(10^{-5}k_{\rm i})\sim c$ is generated, starting from the initial condition $\zeta(k_{\rm i})=0$, cf.~Fig.~\ref{fig: RegenFLow}.
\begin{figure}[t]
	\centering
	\includegraphics[width=\linewidth]{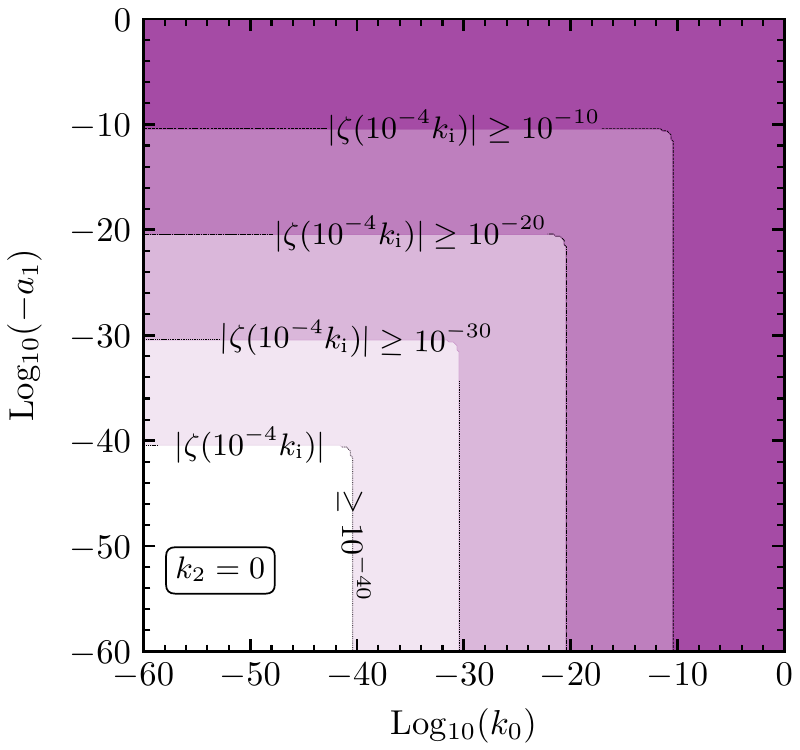}
	\caption{For $\lambda=0$, $g=1$ and $k_2=0$, we show the value of~$\zeta(10^{-4}k_{\rm i})$ generated by the RG flow, starting from the initial condition $\zeta(k_{\rm i})=0$ at the transplanckian UV scale $k_{\rm i}$. The colored regions indicate where~$\zeta(10^{-4}k_{\rm i})$ exceeds a certain value.  }
	\label{fig: k20exclFlow}
\end{figure}
\begin{figure}[t!]
	\centering
	\includegraphics[width=\linewidth]{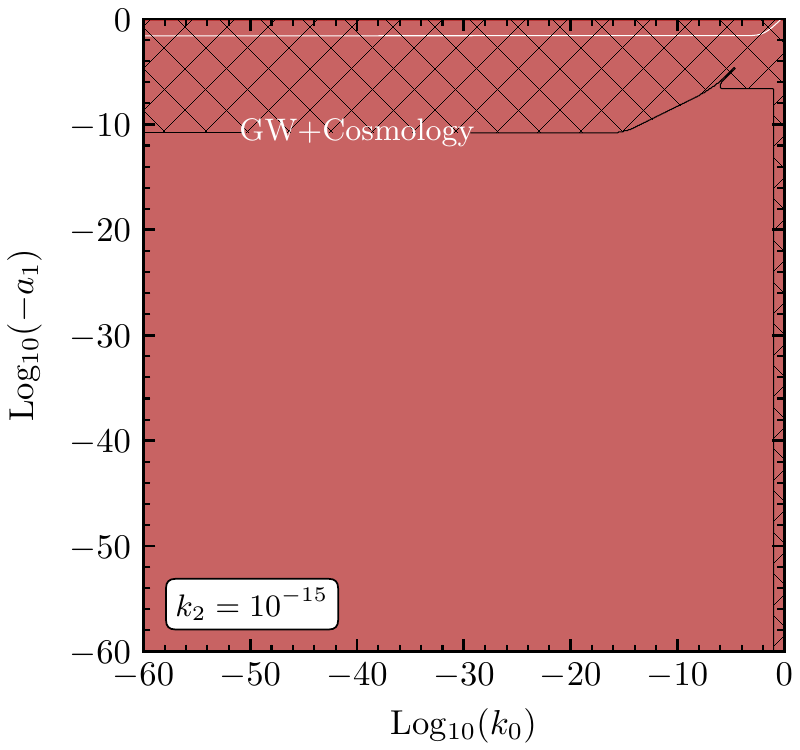}
	\caption{Exclusion plot for $\lambda<\lambda_{\rm crit}$ with initial conditions close to the IR repulsive interacting fixed point $\zeta_{*,2}$. We display the results for the specific choice $\lambda=-11/2$ (which is generic for the purposes of this argument). The red region marks the area where $|\zeta|>10^{-10}$. The white line is the only allowed region: it is a band with width of twice the bound,~$2\,\zeta_{\rm exp}$, centered on values of~$a_1$ and~$k_0$ that render $\zeta_{*,2}$ exactly zero, cf.~Eq.~\eqref{eq: NGFPIRREP}. The white band lies  within the black hatched region, that marks the range of values in the $(a_1,k_0)$--plane already excluded by direct observations \cite{Gumrukcuoglu:2017ijh}.}
	\label{fig: ExclPlotPosNGFP}
\end{figure}
\begin{figure}[t!]
	\centering
	\includegraphics[width=\linewidth]{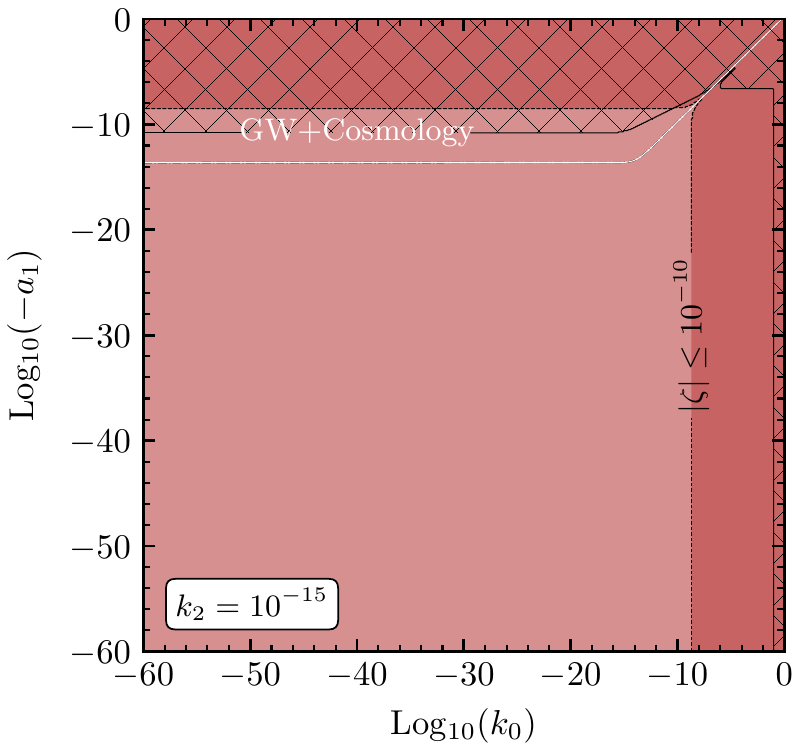}
	\caption{Exclusion for~$\lambda<\lambda_{\rm crit}$ and~$k_2=10^{-15}$ with initial conditions in the basin of attraction of the IR attractive sGFP, cf.~\eqref{eq: sGFPbelow}. The dark and light red regions indicate the areas where $\zeta_{*,1}>10^{-20}$ and $\zeta_{*,1}>10^{-10}$, respectively. The  white line corresponds to $\zeta_{*,1}=0$, while the black hatched region indicates the area of exclusion by direct observations~\cite{Gumrukcuoglu:2017ijh}.}
	\label{fig: k20exclreg}
\end{figure}
\begin{figure*}[t!]
	\includegraphics[width=\linewidth]{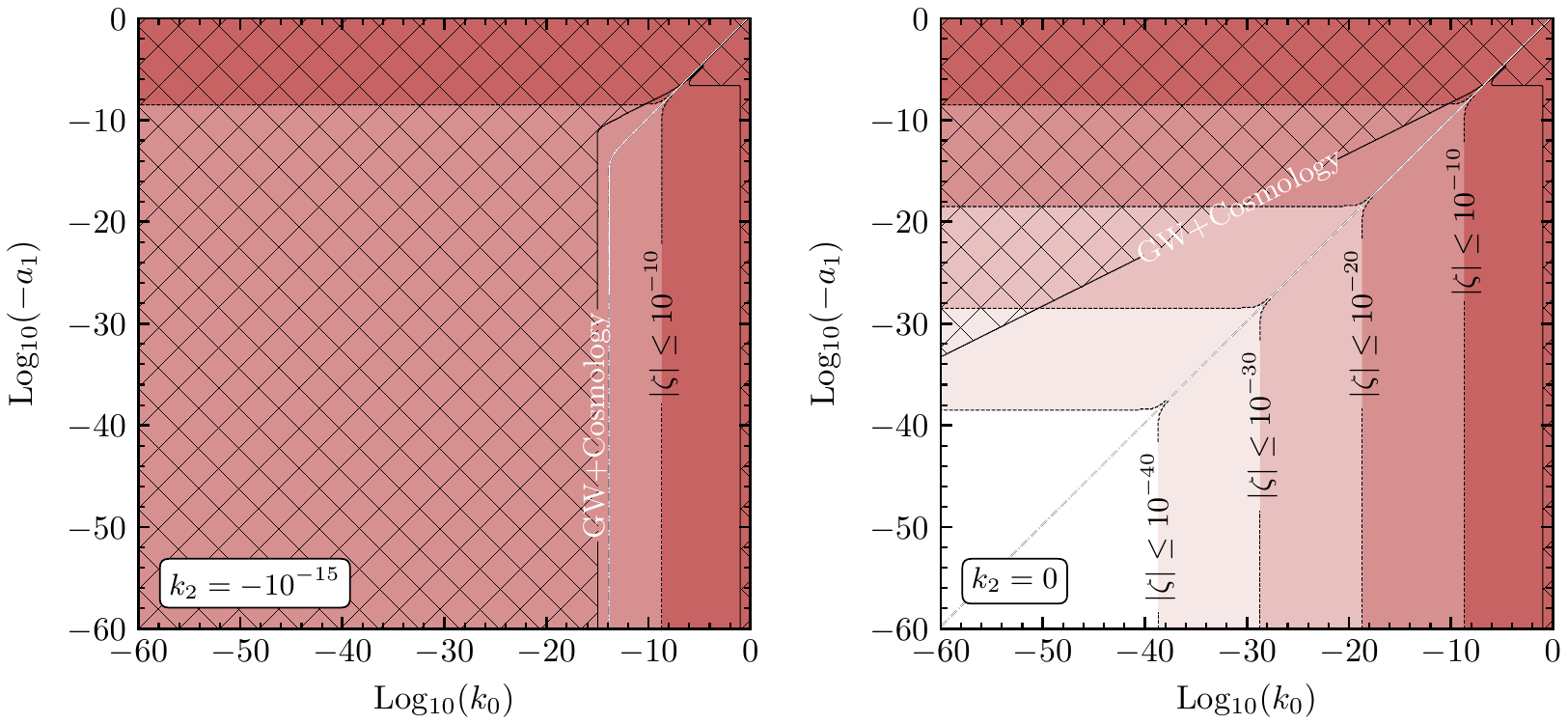}
	\caption{ 
		Exclusion for $\lambda<\lambda_{\rm crit}$ with initial conditions in the basin of attraction of the IR attractive sGFP, cf.~\eqref{eq: sGFPbelow}. The red region marks the area where the value of~$\zeta_{*,1}$ exceeds a certain bound, e.g., $|\zeta|\leq10^{-10}$. The different contours and shades exclude the corresponding area for different values of the assumed  bound. The white line indicates $\zeta_{*,1}=0$, and the hatched region marks area of exclusion by direct observations~\cite{Gumrukcuoglu:2017ijh}. The left panel shows the case $k_2=-10^{-15}$, while the right panel refers to the case $k_2=0$.}
	\label{fig: k2posexclreg}
\end{figure*}
This behavior follows from the dependence of the sGFP~$\zeta_{*,1}$ on the gravitational LIV couplings.
For the case $\lambda=0$ (which is generic for the purposes of this argument), it reads
\begin{equation}
\zeta_{*,1}=\frac{-42 a_1+43 k_0-113 k_2}{2016}\,,
\end{equation}
such that very small or very special values for the gravity LIV couplings are necessary to accommodate a small value for the sGFP.  From Fig.~\ref{fig: k20exclFlow}, we can  estimate how non-generically the initial conditions have to be chosen above the Planck scale to accommodate the strong bounds at lower scales: Starting from $\zeta(k_{i})=0$, a flow over
four orders of magnitude is sufficient to regenerate a non-vanishing value of $\zeta$. Imposing constraints on the IR value of $\zeta$ as they arise from the corresponding observations on $\zeta$ in the full Standard Model would result in the conclusion that these initial conditions are excluded, except for very special points in the gravitational parameter space, i.e., exactly on the fixed point itself or tiny deviations around it.\\

\paragraph{Constraints on LIV gravity couplings for $\lambda<\lambda_{\rm crit}$}\textcolor{white}{b}\\
We now focus on the case of $\lambda<-\tfrac{21}{4}$. We emphasize that $\lambda$ pertains to the microscopic value of the dimensionless cosmological constant.
Quantum fluctuations of matter might drive the cosmological constant to positive values in agreement with observations in the IR, starting from initial conditions at negative $\lambda$ in the UV~\cite{Dona:2013qba}. 
For a specific realization in the case of Lorentz-invariant gravity, see~Fig.~4 in~\cite{deAlwis:2019aud}, where one explicit trajectory in the approximation of~\cite{Dona:2013qba} was employed.

For $\lambda<\lambda_{\rm crit}$, the interacting fixed point~$\zeta_{*,2}$ is IR repulsive and therefore shields all initial conditions with~$\zeta(k_{\rm i})<\zeta_{*,2}$ from a phenomenologically viable regime. Specifically, small deviations from~$\zeta_{\ast,2}$ at~$k_{\rm i}$ will increase towards lower scales. In a similar manner to the case analyzed in the previous section, this results in strong constraints on the gravitational LIV couplings.
For the case of initial conditions exactly on the fixed point, a small value of~$\zeta(M_{\rm Pl})$ can still be achieved. The white line in Fig.~\ref{fig: ExclPlotPosNGFP} shows where~$\zeta(M_{\rm Pl})=0$ can be satisfied. The allowed region for~$|\zeta_{*,\,2}|<\zeta_{\rm exp}$ with some experimental bound~$\zeta_{\rm exp}$ corresponds to a band with the width of~$2\,\zeta_{\rm exp}$ around the white line. As can be understood from the linear expansion of this fixed point for the generic choice  of $\lambda=-11/2$,
\begin{equation}
\label{eq: NGFPIRREP}
\zeta_{*,2}\approx-0.317 a_1-0.597 k_0-7.518 k_2-0.00792\,,
\end{equation}
this cancellation can only happen for~$-a_1 \sim \mathcal{O}(10^{-2})$, which is excluded by observations~\cite{Gumrukcuoglu:2017ijh}. 
For initial conditions with~$\zeta(k_{\rm i})>\zeta_{*,2}$, the RG trajectories are focused by the IR attractive sGFP $\zeta_{*,1}$ towards a universal value~$\zeta(M_{\rm Pl})\approx\zeta_{*,1}(M_{\rm Pl})$.  This universal value depends on the LIV couplings~$k_0,\,k_2$ and~$a_1$.
In turn, this allows to translate the constraints on~$\zeta(M_{\rm Pl})$ into strong constraints on~$k_0,\,k_2$ and~$a_1$. 
We use the observational constraints on~$\zeta(0)$ to constrain~$\zeta(M_{\rm Pl})$ in our toy model. With the corresponding caveats in mind, we extract bounds on the gravitational LIV couplings, as shown in the exclusion plots~Figs.~\ref{fig: k20exclreg} and~\ref{fig: k2posexclreg}. There, we 
combine the constraints on the gravity LIV couplings coming from the strong constraints on~$\zeta_{*,1}$ with the existing constraints from cosmology and the observation of gravitational waves from a neutron-star merger with electromagnetic counterpart \cite{TheLIGOScientific:2017qsa,GBM:2017lvd,Monitor:2017mdv}. We focus on the  $(k_0,a_1)$--plane for different values of $k_2$, since the observation of gravitational waves leads to the strong constraint~$|k_2|<10^{-15}$~\cite{Gumrukcuoglu:2017ijh}\footnote{Strictly speaking this bound is obtained by the LIGO data \cite{Monitor:2017mdv} assuming that the speed of photons remains unchanged, i.e. $v_{\gamma}=1$. Neglecting the difference between the Abelian gauge field $A_{\mu}$ in our work and photons, in the present context, the photons are expected to propagate with $v_{\gamma}=1+C\,\zeta$, $C$ being a constant, due to the presence of the LIV coupling $\zeta$. Therefore the observation of gravitational waves from a neutron-star merger with electromagnetic counterpart leads to a constraint $|k_2-C\,\zeta|<10^{-15}$. While for $k_2=0$ this would actually constrain the value of $\zeta$, we do not use this constraint, to emphasize the difference between our toy model containing the Abelian hypercharge, and the measurement involving the photon.}.

For the generic choice of $\lambda=-11/2$, the IR attractive sGFP to linear order in the gravity LIV couplings reads
\begin{equation}
\label{eq: sGFPbelow}
\zeta_{*,1}=\frac{186 a_1+325 k_0+4297 k_2}{576}\,.
\end{equation}
Hence, for~$k_2=\pm 10^{-15}$, the viable region with~$|\zeta_{*,1}|<\zeta_{\rm exp}$, where~$\zeta_{\rm exp}$ is the experimental bound, is a band with width of~$2\zeta_{\rm exp}$ in the $(a_1,k_0)$--plane,~cf.~Figs.~\ref{fig: k20exclreg} and~\ref{fig: k2posexclreg}. We emphasize that the fixed point $\zeta_{*,1}$, which leads to a universal value of $\zeta$ at the Planck scale, goes over into the GFP in the limit of vanishing LIV couplings. Therefore, its existence is expected also beyond the present truncation, such that the qualitative features of the above analysis are expected to carry over to extended truncations. Furthermore, if we assume the interacting fixed point $\zeta_{*,2}$ to be an artifact of the truncation, the region~$\zeta(k_{\rm i})<\zeta_{*,2}$ is not shielded from the phenomenologically viable region. Consequently, any value of $\zeta(k)$ would lie in the basin of attraction of the IR attractive FP $\zeta_{*,1}$, leading to constraints on the gravity LIV couplings.

\section{Modified dispersion relations}
\label{sec: HigherOrders}
The experimental study of Lorentz symmetry violation often proceeds by constraining each term in the SME separately. Within a given theoretical setting, these terms are typically not independent. This is important in light of the fact that leading order, marginal couplings are typically much simpler to constrain experimentally. In contrast, higher-order terms are generically Planck-scale suppressed, and therefore hard to strongly constrain by observations. Yet, within a given theoretical setting, consistency conditions link these couplings. These conditions can be derived from their beta functions. To exemplify this idea, let us explore terms which modify the dispersion relation for propagating Abelian gauge fields.
Due to their canonical mass dimension, the corresponding couplings are expected to feature an IR-attractive (shifted) Gaussian fixed point. Accordingly, their IR value is a universal prediction of the theory in the same way as exploited in the previous sections. 

As an example, we consider the higher-order operator~$\bar{\kappa}\,\,n_{\alpha} n_{\beta}D^{\alpha}D^{\beta}\,F^{\mu\nu}F_{\mu\nu}$. Such a term gives rise to a higher-order dependence on the energy in the dispersion relation, i.e., 
\begin{equation}
\vec{p}^2 = E^2 + \frac{\kappa}{M_{\rm Pl}^2}E^4.
\end{equation}
 For photons, such modifications have received significant observational interest, see, e.g., \cite{Jacobson:2001tu,Jacobson:2002ye,Bolmont:2006kk,Ackermann:2009aa,Vasileiou:2013vra,Ellis:2018lca,Abdalla:2019krx} and references therein.
Within our setup, the cubic term, explored in an EFT setting, e.g., in \cite{Myers:2003fd}, can actually be set to zero consistently. We focus on the dimensionless counterpart of the coupling $\bar{\kappa}$, i.e., $\kappa = \bar{\kappa}/k^2$.
Instead of considering the full beta function for $\kappa$, we limit our study to the \emph{inducing} term,  
 i.e., the analog of $b_0$ in Eq.~\eqref{eq: betaschem}. It reads
\bea
b_{0,\,\kappa}&=& g\,\bigg(\frac{815 a_1-179 k_0-1847 k_2}{1080 \pi  (1-2 \lambda )^2}\notag\\
&{}&\hspace{18pt}+\frac{230 a_1+262 k_0+1021 k_2}{540 \pi  (1-2 \lambda )^3}\bigg)\notag\\
&{}&+\zeta\,  g\, \bigg(-\frac{4}{3 \pi  (1-2 \lambda )}+\frac{1685 a_1-1905 k_0+1204 k_2}{2160 \pi  (1-2 \lambda )^3}\notag\\
&{}&\hspace{28pt}-\frac{3167 a_1-243 k_0+280 k_2+5760}{2160 \pi  (1-2 \lambda )^2}\bigg)\notag\\
&{}&+\zeta ^2\, g\,\bigg(
\frac{9}{4 \pi  (1-2 \lambda )}+\frac{-17277 a_1+6809 k_0+9807 k_2}{5400 \pi  (1-2 \lambda )^3}\notag\\
&{}&\hspace{27pt}+\frac{24819 a_1+3797 k_0-5109 k_2+32580}{10800 \pi  (1-2 \lambda )^2}\bigg).
\label{eq: b0HO}
\eea
The first line shows that $\kappa$ is induced by gravitational fluctuations in the presence of the LIV couplings $k_2, k_0, a_1$. The following lines highlight that $\zeta$ is also induced once a finite $\zeta$ is present. Lorentz symmetry breaking therefore percolates from the gravitational to the matter sector, but also spreads within the matter sector, once a ``seed" in the form of one nonvanishing LIV matter coupling is present.

Including the term linear in $\kappa$, which also accounts for the canonical mass dimension of $\kappa$, the beta function reads
\begin{equation}
	\beta_{\kappa}=b_{0,\,\kappa}+2\,\kappa+b_{1,\,\kappa}\,\kappa\,,
	\label{eq: betaHO}
\end{equation} 
where the second term is the contribution due to the canonical mass dimension of $\bar{\kappa}$.
From Eq.~\eqref{eq: betaHO}, the fixed-point value of $\kappa$ is given by
\be \label{eq: kappastar}
\kappa_{*} = -\frac{b_{0,\, \kappa}}{2+b_{1,\,\kappa}}.
\ee
This relation holds under the self-consistency assumption that~$\kappa_{\ast}\ll1$, as then terms quadratic in $\kappa$  can be neglected in $\beta_{\kappa}$. The critical exponent of this fixed point~\eqref{eq: kappastar}~is $\theta=-2-b_{1,\,\kappa}$. This fixed point is IR attractive, as long as metric fluctuations remain near-perturbative, i.e., $|b_{1,\,\kappa}|<1$. 
Due to the constant offset in the denominator, the fixed-point value $\kappa_*$ is parametrically set by the fixed-point value for $\zeta$ and the values of~$k_0, \,k_2$ and~$a_1$. Following the same logic as in the previous section, the value of $\kappa$ at the Planck scale corresponds to the fixed-point value~$\kappa_\ast$. Below the Planck scale, gravitational fluctuations turn off dynamically, resulting in a vanishing flow for $\kappa$ except for the dimensional term. This has the simple solution
\be
\kappa(k<M_{\rm Pl}) = \kappa_{*} \left(\frac{k}{M_{\rm Pl}}\right)^2.
\ee
For the dimensionful counterpart $\bar{\kappa}$, this implies
\be
\bar{\kappa}(k<M_{\rm Pl}) = \frac{\kappa_{*}}{M_{\rm Pl}^2}.
\ee
Let us briefly compare this with experimental constraints on the quartic term in the dispersion relation for photons, which constrains the dimensionless coupling to be $|\kappa_{\rm exp}|<10^6$ \cite{Kostelecky:2013rv},  see also \cite{Vasileiou:2013vra}. 
In contrast, a significantly stronger indirect constraint can be obtained by choosing $\zeta,\, k_2,\, k_0$ and~$a_1$ such that they satisfy the corresponding constraints but maximize $b_{0\, \kappa}$. With the exception of very special points in the parameter space, this generically constrains $b_{0\, \kappa}$ to about the same order as $\zeta$ itself. Conversely, within a setting described by our toy model we would not expect direct searches for $\kappa$ to result in a detection, unless a rather significant improvement was achieved in future observations. We emphasize that the above analysis, especially the restriction to the inducing contribution to $\beta_{\kappa}$, is only the first step in the analysis of effects of the higher order coupling $\kappa$. However, the qualitative feature of the sGFP $\kappa_{*}$, i.e., the form of Eq.~\eqref{eq: kappastar}, will remain unchanged under the consideration of the full $\beta_{\kappa}$, since any direct contribution will contribute to $b_{1,\, \kappa}$ or $b_{2,\, \kappa}$. Therefore, the qualitative feature that $\kappa_*$ serves as an IR attractor, with the value parametrically set by $\zeta_*$, already follows from the analysis of the inducing term $b_{0,\, \kappa}$.

\section{Conclusions and outlook}
\label{sec: Concl}
Probing the quantum nature of gravity observationally is notoriously difficult. The interplay of quantum gravity with matter could become a key stepping stone for progress in this direction: At the microscopic level, this interplay could determine properties of elementary particles that are accessible to experiments at lower energies. Thereby, low-energy (sub-Planckian) measurements could constrain transplanckian Physics. 
This idea to use matter as a ``magnifying glass" for the quantum properties of spacetime underlies part of the swampland-program in string theory, as well as a similar program within the asymptotic-safety approach. Here, we highlight the potential power of such considerations for Lorentz-invariance violating gravity-matter models.

The key idea underlying this paper is the following: Quantum fluctuations of gravity that only respect foliation-preserving diffeomorphisms generate Lorentz-invariance violating interactions for matter, in our case parameterized by the scale dependent coupling $\zeta(k)$. Within the toy model we consider and within our approximation of the dynamics,
the corresponding beta function features an infrared-attractive fixed point. Its value is determined by the gravity-LIV couplings. Under appropriate conditions, spelled out in this paper, it governs the scale dependence of $\zeta$.
Due to its infrared-attractive nature, it results in a \emph{universal} value of~$\zeta$ at the Planck scale, which is independent of the initial conditions for~$\zeta(k_i)$ at the high-energy scale $k_i$, but depends on the values of the gravity-LIV couplings. The Planck-scale value of $\zeta$ can be mapped to its low-energy value by the standard RG flow without gravity. At low energies, experimental constraints on LIV-matter couplings exist. Such low-energy experimental bounds also indirectly constrain the values attained by the LIV-matter couplings at Planckian scales. As the latter Planck-scale values depend on the LIV gravitational couplings, experimental bounds on LIV-matter couplings can be translated into constraints on the LIV couplings of the gravitational sector. Moreover, as observational constraints on matter-LIV couplings are rather strong for marginal couplings, this mechanism can provide constraints on the gravity-LIV couplings which are significantly stronger than the direct observational constraints.

To support this general idea, we have performed a study of an Abelian gauge field coupled to gravity with foliation-preserving diffeomorphisms only. Our study has the following technical limitations: It is performed in a truncation of the full dynamics, as the RG flow generates additional terms. We do not account for their feedback. This results in a systematic uncertainty of our results. Further, we work in a Euclidean setup in order to apply FRG techniques.  The presence of a  foliation should ensure that a Wick-rotation exists. Finally, we work in a toy model for the photon-gravity system: We do not account for the additional matter degrees of freedom of the Standard Model, and neglect electroweak symmetry breaking which implies that the U(1) gauge field relevant at high energies is not the same as the photon of electromagnetism. Bearing these limitations in mind, our study supports the general idea explained above.

Specifically, we have shown that the breaking of Lorentz symmetry in the gravitational sector automatically percolates into the matter sector. This result is entailed in the $\zeta$-independent part of the beta function for $\zeta$, cf.~Eq.~\ref{eq: betazeta}. This term measures  the ``amount of LIV'' in the gravitational sector that impacts the matter sector, and it vanishes if gravity retains full diffeomorphism invariance. Due to this term, $\zeta=0$ is no longer a fixed point of the RG flow. Thus, a non-vanishing $\zeta$ is generated by the flow, even if it is set to zero at some initial scale. Consequently, Lorentz symmetry violation necessarily percolates from the gravitational to the matter sector.

Furthermore, as we have shown within our approximation, the beta function for~$\zeta$ always features an IR-attractive fixed point, that can be reached from a wide range of initial conditions in the far UV, i.e., at transplanckian scales. Therefore, we can remain agnostic about the ultimate UV completion of the theory: As long as it sets initial conditions for the couplings within the appropriate range, there will be a \emph{universal} Planck-scale value of~$\zeta$, corresponding to the IR-attractive fixed point of its RG flow.
In this case, the value of $\zeta$ at the Planck scale is fully determined by the values of the LIV couplings in the gravitational sector, i.e., $k_0, k_2$ and $a_1$ (cf.~Eq.~\eqref{eq: breakingaction}), as well as the dimensionless cosmological constant $\lambda$. The RG flow below the Planck scale is trivial in our setting, where gravitational fluctuations switch off dynamically, resulting in $\zeta(k=0)\approx \zeta(M_{\rm Pl}).$ 

To exemplify the constraining power of such a fixed point, we translate the stringent experimental bounds on the actual photon-LIV coupling, cf.~Tab.~\ref{tab: LIVMatterConstraints}, into bounds on the gravity LIV couplings $k_0,k_2$ and $a_1$ using the fixed-point relation. Note that for quantitatively robust constraints, this should be repeated in an extended study accounting for the presence of additional degrees of freedom and reducing systematic uncertainties by working within extended truncations. We highlight that if we nevertheless used the fixed-point relation for $\zeta$ that arises from our calculation, even the least stringent observational bound on $\zeta$ would exclude an additional area in the parameter space spanned by the gravity-LIV coupling that is not excluded by the observation of gravitational waves, the BBN and ppN constraints. This highlights the power of an IR-attractive fixed point which is related to the breaking of some symmetry: If this symmetry-breaking is strongly constrained in one sector of the system, an IR-attractive fixed point in this sector can be used to translate observational bounds into constraints on the other sector. A future analysis of the RG flow of the combined gravity-matter system, including the scale dependence of the gravity LIV couplings, would allow to identify intervals of initial conditions for the gravitational LIV couplings, for which the scenario presented in this paper is valid.

Finally, we have also shown that  a higher-order LIV coupling $\kappa$ -- related to a modification of the dispersion relation -- is induced in the same way as $\zeta$ is induced. Due to its canonical mass dimension, it is expected to feature an IR attractive fixed point, whose value is parametrically of the same size as the fixed-point value of $\zeta$. Therefore, constraints on $\zeta$ restrict the possible values of $\kappa_*$. Due to their Planck-scale suppression, this kind of irrelevant coupling is experimentally  less strongly constrained. We therefore estimate that the parametric dependence of $\kappa(M_{\rm Pl})$ on $\zeta(M_{\rm Pl})$, together with the strong constraints on $\zeta$ could result in strong, indirect constraints of $\kappa$. A more extended analysis including the entire beta-function of $\kappa$ and an analysis of the fixed-point structure can confirm this expectation.\\

\emph{Acknowledgements:} We thank B.~Knorr and S.~Lippoldt for insightful discussions. This work is supported by the DFG under grant no.~Ei-1037/1, and A.~E.~is also partially supported by a visiting fellowship at the Perimeter Institute for Theoretical Physics. This
  research is also supported by the Danish National Research
  Foundation under grant DNRF:90. A.~P.~is supported by the Alexander von Humboldt Foundation and M.~S.~is supported by a scholarship of the German Academic Scholarship Foundation. A.~E.~would like to acknowledge the contribution of the COST Action CA18108: QG-MM (Quantum Gravity phenomenology in the Multi-Messenger approach). M.S.~gratefully acknowledges the hospitality at CP3-Origins during the final stages of this work.

\appendix
\section{Foliated spacetimes and the functional Renormalization Group}
\label{sec: AppFRG}
\subsection{Functional Renormalization Group setup for the metric-matter system}
\label{sec: FRG}

The system we study comprises gravity as well as an Abelian gauge field. To investigate the RG flow of this system, we employ the functional Renormalization Group (FRG). This is based on a scale-dependent effective action $\Gamma_k$, whose flow is given by the Wetterich equation~\cite{Wetterich:1992yh, Ellwanger:1993mw,Morris:1993qb,Reuter:1993kw,Tetradis:1993ts},
\begin{align} \label{eq: flow}
k\partial_{k}\Gamma_k [\Phi ; \bar{g}] = 
\frac{1}{2} {\rm STr} \left[
\big( \Gamma_k^{(2)}[\Phi ; \bar{g}] + R_{k}[\bar{g}] \big)^{-1} \dot{R}_{k}[\bar{g}] \right].
\end{align}
By $\Gamma_k^{(2)}$ we denote the second functional derivative of $\Gamma_k$ with respect to the fields.
Here, the superfield $\Phi$ is a collection of all dynamical fields, which in the context of the present work will be
\begin{equation}
\Phi^{\rm A}=(h_{\mu\nu},n_{\mu},A_{\mu}).
\end{equation}
Here, $h_{\mu\nu}$ are metric fluctuations, $n_{\mu}$ is a normalized vector that singles out a preferred frame, and $A_{\mu}$ is the Abelian gauge field. The super-trace $\rm STr$ includes a summation over all discrete indices and an integration over continuous  coordinates. Further, it also implements a trace in field space, since $\Gamma_k^{(2)}$ is actually a matrix in field space, spanned by the inverse propagators, as well as mixed entries.
Finally, the function $R_{\rm k}$ is a scale-dependent infrared regulator, which implements the Wilsonian idea of momentum-shell wise integration of quantum fluctuations, and ensures the UV and IR finiteness of the Wetterich equation. 
In particular, throughout this work we will employ a Litim-type regulator~\cite{Litim:2001up}. 
{For gravity,}  the cutoff function~$R_k$ has to be set up with respect to a background metric~$\bar{g}$ in order to allow for a local coarse graining and the definition of a momentum. As a consequence, the regulator term is, besides the gauge-fixing action, a second source of breaking of diffeomorphism invariance. 

We highlight a key advantage of the Wetterich equation: It depends on the full, non-perturbative (and field dependent) propagator~$(\Gamma_k^{(2)}+R_k)^{-1}$, but is structurally a one-loop equation. This is central to allow feasible calculations in a gravitational context. In the derivation of the Wetterich equation, this one-loop structure is ensured by introducing the regulator as a mass-like term, i.e., introducing the quadratic interaction $\Phi R_k \Phi$ into the generating functional. 
In addition, the presence of a background field allows to gauge-fix the gravitational fluctuations. Just as for gauge theories, using the background field method for the gauge fixing allows to preserve a background gauge symmetry. Unlike in the functional quantization of gauge theories, the introduction of a background field is however not optional in a local formulation of gravitational fluctuations. The background field is an auxiliary field and in principle could be kept arbitrary -- in fact, keeping track of the physical metric and the background metric is key to restore background independence \cite{Becker:2014qya}  -- but specific choices greatly simplify the calculations. In particular, for the projection on curvature-independent matter interactions, the choice of background does not matter. A flat background is the technically simplest choice in this case. In the following, we will therefore adopt this strategy and choose a flat background metric, i.e.,
\begin{equation}\label{eq:BGmetric}
\bar{g}_{\mu\nu}=\delta_{\mu\nu}.
\end{equation}
The Wetterich equation leads to a tower of coupled differential equations that encode the scale-dependence of all  infinitely many couplings of the theory space. In practice, this tower has to be restricted to a, typically finite, subset of equations. Therefore, all results are subject to systematic errors, which have to be estimated by studies of residual gauge, regulator dependences and changes under the extension of the truncation.  To set up our truncation, we choose an ansatz for the scale dependent effective action $\Gamma_k$, which is expanded in terms of metric fluctuations $h_{\mu\nu}$ around the background metric~\eqref{eq:BGmetric},
\begin{equation}
h_{\mu\nu}=g_{\mu\nu}-\delta_{\mu\nu}.
\label{eq: split}
\end{equation}
Since in the present work, we will investigate the effect of metric fluctuations on matter couplings, an expansion up to second order in metric fluctuations is sufficient.
The ansatz for the Lorentz invariant part for the present metric-matter system is given by
\begin{align}
\Gamma^{\rm LI}_k&=\Gamma_k^\mathrm{Abelian}+\Gamma_k^\mathrm{Abelian,\,gf}+\Gamma_k^\mathrm{EH}+\Gamma_k^\mathrm{Grav,\,gf}\,,
\end{align}
cf.~Eqs.~\eqref{eq: EH}, \eqref{eq:gaugekinetic}. We work with the standard gauge-fixing term for the Abelian gauge field,
\begin{equation}
\Gamma_k^{\rm Abelian, \, gf}= \frac{1}{\xi} \int \!\! \mathrm{d}^{4} x \sqrt{\det(\bar{g}_{\kappa\epsilon})} \, (\bar{g}^{\mu\nu}\bar{D}_{\mu}A_{\nu})^2,
\quad \xi \to 0,
\end{equation} 
and for gravity,
\begin{equation}
\Gamma_k^{\rm Grav, \, gf}
=  \int \!\! \mathrm{d}^{4} x \frac{\sqrt{\det(\bar{g}_{\kappa\epsilon})}}{32 \pi {G}_{\rm N}(k) \, \alpha} \, F^{\mu} \bar{g}_{\mu\nu} F^{\nu},
\quad \alpha \to 0,
\end{equation}
with the gauge-fixing condition
\begin{equation}
F^{\mu}
= \left( \bar{g}^{\mu \kappa} \bar{D}^{\lambda}
- \frac{1+\beta}{4} \bar{g}^{\kappa\lambda} \bar{D}^{\mu} \right) h_{\kappa\lambda},
\qquad \beta = 1.
\end{equation}
Note that this gauge fixing also gives rise to Fadeev-Popov ghosts in the gravitational sector. However, since in this work we neglect induced ghost-matter interactions, they do not contribute in the present computations. The Fadeev-Popov ghosts for the Abelian gauge sector decouple from the Abelian gauge field, but contribute to the running of gravitational couplings. As these are not studied in this work, the Abelian Fadeev-Popov ghosts can be neglected.

Finally, despite the breaking of diffeomorphism invariance due to gauge fixing and regulator, we will assume that all different $n$-point functions originating from the same term in the effective  action $\Gamma_k$ come with the same coupling. Indications that this assumption holds semi-quantitatively at an asymptotically safe fixed point has been investigated in a Lorentz invariant setting, e.g., in \cite{Denz:2016qks,Eichhorn:2018akn,Eichhorn:2018ydy,Eichhorn:2018nda}.

\subsection{Functional Renormalization Group setup for the foliation structure and Lorentz-symmetry violations}
\label{sec: Foliation}

In order to study the effect of operators which are invariant under foliation preserving diffeomorphisms, as a first step the implementation of a foliation structure on the four dimensional Euclidean space(time) is necessary.
In order to implement a foliation structure, and restrict the sum over all metrics in the gravitational functional integral to a sum over globally hyperbolic spacetimes, we need to resort to a suitable parameterization for the full metric and its fluctuations.

One common choice of parametrization in the context of foliated spacetimes and Lorentz-symmetry breaking theories is to express the full metric in terms of ADM variables~\cite{Arnowitt:1959ah,Arnowitt:1962hi,Manrique:2011jc,Rechenberger:2012dt,Biemans:2016rvp,Biemans:2017zca,Platania:2017djo,Houthoff:2017oam}. To parametrize fluctuations, each of these fields is split linearly into background and fluctuation quantities, i.e., in an analogous way to Eq.~\eqref{eq: split}.  The advantage of this procedure is that the choice of ADM variables automatically ensures the foliation structure of the full metric. However, the map between the metric fluctuations~$h_{\mu\nu}$ and the fluctuations of the ADM fields is non-linear.

As pointed out in the previous section, \emph{iff} the regulator term is quadratic in the fluctuation fields, the flow equation~\eqref{eq: flow} is structurally a one-loop equation. If we imagine the path integral for gravity to be defined in terms of the fluctuation field $h_{\mu\nu}$, with an appropriate quadratic regulator, a transition to ADM variables is disastrous: Due to the non-linearity of the metric fluctuations~$h$ in terms of the fluctuations of the ADM fields, implementing a foliation structure via ADM variables would break the one-loop structure of the Wetterich equation. 
Preserving the one-loop structure of the flow equation within the ADM setup requires to define the regularized path integral directly at the level of the ADM fields, such that the regulator term $\Phi R_k \Phi$ is quadratic in the fluctuations of the ADM fields.

In addition, we require the regulator term to be invariant under the auxiliary background gauge invariance. Yet, for full diffeomorphism invariance, the ADM fluctuation fields transform non-linearly under the gauge transformation. Therefore, a regulator term quadratic in the ADM-fluctuations fields would break full background diffeomorphim invariance down to foliation preserving diffeomorphims. In other words, within the ADM formalism, writing a quadratic regulator that preserves background gauge invariance while arising from a linear split of the original metric into background and fluctuation, appears impossible, as emphasized in \cite{Rechenberger:2012dt}. In contrast, in a setting with foliation-preserving diffeomorphisms, the symmetry acts linearly on the ADM fields, allowing a standard construction of a flow equation. However, the main purpose of our work it to understand whether and how LIV-terms in the matter sector can be induced by quantum gravitational fluctuations in the presence of LIV gravitational couplings. To this end, it is crucial to have a subsector of the gravity theory (parameterized by the Einstein-Hilbert action) that preserves full diffeomorphism symmetry. In this way, we can cleanly distinguish that it is the LIV-terms in the gravitational sector that induce LIV terms in the matter sector, and that it is not gravitational fluctuations \emph{per se} which result in LIV terms in the matter sector.

We will therefore employ an alternative formalism proposed in \cite{Knorr:2018fdu}, which ensures the required symmetries. In this approach, the full metric $g_{\mu\nu}$ is written in terms of a spatial metric $\sigma_{\mu\nu}$ and a normalized time-like vector $n_\mu$, according to Eq.~\eqref{eq: cond}. On the level of the fields $\sigma_{\mu\nu}$ and $n_{\nu}$, the split~\eqref{eq: split} of the full metric into background and fluctuations is parametrized by
\bea
n_{\mu}&=&\bar{n}_{\mu}+\hat{n}_{\mu},\notag\\
\sigma_{\mu\nu}&=&\bar{\sigma}_{\mu\nu}+\hat{\sigma}_{\mu\nu}-\hat{n}_{\mu}\hat{n}_{\nu} .
\label{eq: flucts}
\eea
Due to the nonlinear split of $\sigma_{\mu\nu}$, this amounts to a linear parametrization of the metric fluctuations in terms of the fluctuations of $n_{\mu}$ and $\sigma_{\mu\nu}$
\begin{equation}
h_{\mu\nu}=\hat{\sigma}_{\mu\nu}+\bar{n}_{\mu}\hat{n}_{\nu}+\hat{n}_{\mu}\bar{n}_{\nu}.
\label{eq: trafo}
\end{equation}
The linearity of $h_{\mu\nu}$ in the fluctuating fields $(\hat{n},\hat{\sigma}_{\mu\nu})$ is the key ingredient for constructing a background-diffeomorphism-invariant flow equation on foliated spacetimes, while preserving the typical one-loop structure of the flow equation in the metric formalism \cite{Knorr:2018fdu}.

The path integral is then restricted to foliated spacetimes by translating the conditions~\eqref{eq: cond} into constraints for  the fluctuation fields~$(\hat{n},\hat{\sigma}_{\mu\nu})$. Both conditions are solved by 
\begin{equation}
\mathcal{F}_{\mu}=\bar{n}^{\nu}\hat{\sigma}_{\mu\nu}-\bar{n}^{\nu}\hat{n}_{\mu}\hat{n}_{\nu}=0.\label{eq: constraint}
\end{equation}
In the present work, this constraint will be implemented akin to a gauge-fixing term, i.e., by introducing and then exponentiating a delta-function of the constraint into the path integral. This procedure results in an additional term, 
\begin{equation}
\Gamma^{\rm Fol}_k=\frac{1}{32\pi G_{\rm N}(k)\alpha_{\rm Fol}}\int\! \sqrt{\det(\bar{g}_{\kappa\epsilon})}\,\bar{g}^{\mu\nu}\,\mathcal{F}_{\mu}\,\mathcal{F}_{\nu},
\quad \alpha_{\rm Fol}\to 0,
\label{eq: SFol}
\end{equation}
into the action. 
Some remarks on Eq.~\eqref{eq: SFol} and the implementation of the second-class constraints \eqref{eq: cond} are necessary. At variance of first-class constraints, i.e. constraints associated with gauge symmetries, second-class constraints cannot be implemented via the Faddeev-Popov trick.
With second-class constraints, a Hamiltonian analysis is actually desirable to understand whether secondary constraints need to be imposed in addition. Within the functional Renormalization Group framework, this is -- to our knowledge -- not well-explored yet.
More details on the implementation of second-class constraints will be discussed elsewhere~\cite{Eichhorn2019}. 

In the present work, we assume that the implementation of the constraint~\eqref{eq: constraint} via the constraint term~\eqref{eq: SFol} is sufficient to capture the relevant features. Additionally, we implement the orthogonality condition Eq.~\eqref{eq: extrorth} of the extrinsic curvature at the level of the  scale dependent effective action $\Gamma_k$.

Employing the parameterization~\eqref{eq: trafo} for the metric fluctuations and implementing the foliation structure via the constraint term~\eqref{eq: constraint}, we can study the effect of operators which are invariant under foliation preserving diffeomorphisms only on the background diffeomorphism invariant system parametrized by~$\Gamma^{\rm LI}_k$. We restrict ourselves to the most relevant operators that break full diffeomorphisms.  Our ansatz for the Lorentz-invariance-violating (LIV) part of the scale dependent effective action is
\begin{equation}
\Gamma_{k}^{\rm LIV}=\Gamma_k^\mathrm{Abelian,\, LIV}+\Gamma_k^\mathrm{Grav,\, LIV}\,,
\end{equation}
with $\Gamma_k^\mathrm{Abelian,\, LIV}$ and $\Gamma_k^\mathrm{Grav,\, LIV}$ specified in Eq.~\eqref{eq: gaugeLIV} and Eq.~\eqref{eq: breakingaction}, respectively.

All other pure-gravity terms which are invariant only under foliation preserving diffeomorphisms containing up to two derivatives, are related via the Gauss-Codazzi equations, up to a total derivative.

For the Abelian gauge field, there are no further non-vanishing invariants at quadratic order in the gauge field:~$n^{\mu}n^{\nu}n^{\kappa}n^{\lambda}F_{\mu\nu}F_{\kappa\lambda}$ vanishes due to the antisymmetry of the field-strength tensor. The operator~$F \tilde{F} = F_{\mu\nu}F_{\kappa \lambda}\epsilon^{\mu\nu\kappa\lambda}$ is a total derivative. As for the invariant $F_{\mu\nu}F_{\kappa\lambda}\epsilon^{\mu\nu\kappa\rho}n_{\rho}n^{\lambda}$, we notice that
\begin{equation}
F_{\mu\nu}\epsilon^{\mu\nu\kappa\rho}F_{\kappa\lambda}\sim(\vec{E}\cdot\vec{B})\delta^{\rho}_{\lambda}.
\end{equation}
Due to the normalization of $n_\mu$, this relation directly leads to
\begin{equation}
F_{\mu\nu}\epsilon^{\mu\nu\kappa\rho}F_{\kappa\lambda}n_{\rho}n^{\lambda}\sim(\vec{E}\cdot\vec{B})\sim F \tilde{F},
\end{equation}
such that also this invariant corresponds to a total derivative and can thereby be neglected. Finally, we explicitly neglect any gauge-symmetry-violating operator generated by the flow, i.e., we work under the assumption that the theory space is spanned by gauge-invariant operators only.
\section{Projection onto the LIV matter coupling}
\label{sec: Porjection}
The flow equation for the system under consideration is obtained by inserting
\begin{equation}
\Gamma_k=\Gamma_k^\mathrm{LI}+\Gamma_k^\mathrm{Fol}+\Gamma_k^\mathrm{LIV}
\end{equation}
into the Wetterich equation~\eqref{eq: flow}. 

In order to derive indirect constraints on the LIV-gravity couplings, based on the experimental and observational bounds on the value of $\zeta$ at low energies, we need to extract the flow of the wave-function renormalization $Z_A$ and the LIV coupling $\zeta$. To that end, we project the RG flow, i.e., the right-hand side of the Wetterich equation, onto the two corresponding invariants $F_{\mu\nu}F^{\mu\nu}$ and $n^{\mu}n^{\kappa}F_{\mu\nu}F_{\kappa}^{\,\,\,\nu}$ (where the appropriate symmetrization is understood implicitly).
We can project the RG flow onto the~$F_{\mu\nu}F^{\mu\nu}$-term by taking two derivatives with respect to~$A_{\mu}$, closing the open indices with a transverse projector, selecting the terms quadratic in external momenta, and subsequently taking the 0th order term in~$n$.  In order to project onto the LIV term~$n^{\mu}n^{\kappa}F_{\mu\nu}F_{\kappa}^{\,\,\,\nu}$, we find that taking two derivatives with respect to the gauge field, closing the indices with a transverse projector, selecting the terms that are quadratic in momenta and taking all terms containing the vector field $n_{\mu}$ or its norm, is already sufficient to isolate this term. 
After this procedure, we set $n^2=1$.

We stress that the projection described here is not unique and that within truncations and due to the breaking of gauge invariance, these different projections might lead to quantitatively different results for the scale dependence of the LIV coupling $\zeta$. 

\bibliography{refs}
\end{document}